\documentclass[12pt]{iopart}

\usepackage{iopams}
\usepackage{amssymb}

\begin{document}

\title[]{Chandrasekhar separation ansatz and the generalized total angular momentum for the Dirac equation in the Kerr-Newman metric.}

\author{D. Batic}

\address{Institute for Theoretical Physics, ETH Z\"{u}rich, CH-8093 Z\"{u}rich, Switzerland}
\ead{batic@itp.phys.ethz.ch}

\author{H. Schmid}

\address{UBH Software \& Engineering GmbH, D-92224 Amberg, Germany}
\ead{Harald.Schmid@UBH.de}

\begin{abstract}
In this paper we compute the square root of the generalized squared total angular momentum operator $J$ for a Dirac particle in the Kerr-Newman metric. The separation constant $\lambda$ arising from the Chandrasekahr separation ansatz turns out to be the eigenvalue of $J$. After proving that $J$ is a symmetry operator, we show the completeness of Chandrasekhar ansatz for the Dirac equation in oblate spheroidal coordinates and derive an explicit formula for the propagator $e^{-itH}$.
\end{abstract}

\ams{83C57, 47B15, 47B25}
\maketitle
\section{Introduction}
One of the most spectacular predictions of general relativity are black holes, which should form when a large mass is concentrated in a sufficiently small volume. The idea of a mass-concentration, which is so dense that even light would be trapped goes back to Laplace in the 18th century. Shortly after Einstein developed general relativity, Karl Schwarzschild discovered in 1916 a mathematical solution to the equations of the theory that describes such an object. It was only much later, with the work of people like Oppenheimer, Volkoff and Snyder in the 1930's, that the scientific community began to think seriously about the possibility that such objects might actually exist in the Universe. It was shown that when a sufficiently massive star runs out of fuel, it is unable to support itself against its own gravitational attraction and it should collapse into a black hole. Only in the 1960's and the 1970's, in the so-called Golden Era of black hole research, new interesting phenomena like the Hawking radiation and superradiance were discovered but for a rigorous mathematical description of them we have to wait until the 1990's and the beginning of the new century, when the rigorous analysis of the propagation and of the scattering properties of classical and quantum fields on black hole space-times was developed.\\
Whenever we attempt to analyze the scattering properties of fields in the more realistic framework of the Kerr-Newman black hole geometry, we are faced with several difficulties, which are not present in the simplified picture of the Schwarzschild metric. First of all, the Kerr-Newman solution is only axisymmetric (cylindrical symmetry), since it possesses only two commuting Killing vector fields, namely the time coordinate vector field $\partial_{t}$ and the longitude coordinate vector field $\partial_{\varphi}$. This implies that there is no decomposition in spin-weighted spherical harmonics and that artificial  long-range terms at infinity will be present in the field equations. Moreover, another difficulty is due to fact that the Kerr-Newman space-time is not stationary. In particular it is impossible to find a Killing vector field which is time-like everywhere outside the black hole. In fact $\partial_{t}$ becomes space-like in the ergosphere, a toroidal region around the horizon. This implies that for field equations describing particles of integral spin (wave equation, Klein-Gordon, Maxwell) there exists no positive definite conserved energy. For field equations describing particles with half integral spin (Weyl, Dirac) we can find a conserved $L^{2}$ norm with the usual interpretation of a conserved charge. Hence, the absence of stationarity in the Kerr-Newman metric is not really a difficulty for the scattering theory of classical Dirac fields.\\
For the reasons mentioned above there are only few analytical studies of the propagation of fields outside a Kerr-Newman black hole. Time-dependent scattering for the Klein-Gordon equation in the Kerr framework has been developed in 2003 by H\"{a}fner \cite{Haf1}, while in 2004 Daud$\acute{\mbox{e}}$ \cite{Da} proved the existence and asymptotic completeness of wave operators, classical at the event horizon and Dollard-modified at infinity, for classical massive Dirac particles in the Kerr-Newman geometry.\\ 
Our present work is the first in a series of three papers devoted to develop a time-dependent scattering theory for massive Dirac particles outside a non-extreme Kerr-Newman background. Our main technical tool will be an integral representation for the Dirac propagator in the non extreme Kerr-Newman metric and asymptotically at infinity, where the Kerr-Newman metric goes over in the Minkowski metric in oblate spheroidal coordinates. The construction of the propagators relies on Chandrasekhar separation of variables, whose completeness plays a crucial role in order to develop a reliable scattering theory. Their integral representations together with estimates for the asymptotic behavior of the solutions of the radial problem, arising in Chandrasekhar ansatz, will allow us to give explicit analytical expressions for wave operators classical at the event horizon and Dollard-modified at infinity. This will be, to our knowledge, the first analytical result, since all previous works in this direction (see \cite{Haf1,Da}), which are based on the Mourre theory, treat only the problem of the existence of the above mentioned wave operators and of the asymptotic completeness for the Dirac equation in the Kerr-Newman metric. Our global strategy is the following:
\begin{itemize}
\item
in the first paper we prove the completeness of Chandrasekhar ansatz for the free Dirac equation in the Minkowski metric in oblate spheroidal coordinates and we derive an integral representation for the Dirac propagator in this framework;
\item 
in the second paper we give an integral representation for the Dirac propagator in the Kerr-Newman geometry in terms of the solutions of the radial and angular ODEs arising in Chandrasekhar separation of variables and we describe the behaviors of the radial solutions close to the event horizon and asymptotically at infinity, respectively; 
\item
in the last paper we treat first classical wave operators asymptotically at infinity and we compute them explicitly by means of the integral representations of the Dirac propagator obtained in the previous two works. In doing so, we get an analytical expression of the time-dependent logarithmic phase shift, that we need in order to construct Dollard-modified wave operators asymptotically away from the black hole. After implementation of this phase shift in the free dynamics we evaluate the Dollard-modified wave operators and obtain for them an integral representation. Then, we compute the wave operators close to the event horizon and prove the asymptotic completeness of the wave operators at the event horizon and asymptotically at infinity.
\end{itemize} 
The present article is structured as follows: In Section 1 we motivate our interest to derive an integral representation for the Dirac propagator in the Minkowski metric in oblate spheroidal coordinates. In Section 2 we introduce the Dirac operator in the Kerr-Newman metric. In Section 3 we find a suitable transformation, which decomposes it  into the sum of an operator containing only derivatives respect to the variables $t$, $r$ and $\varphi$ and of an operator involving only derivatives respect to $t$, $\vartheta$ and $\varphi$. This transformation together with Chandrasekhar ansatz allows us in Section 4 to construct explicitly the generalized total angular momentum $J$ for the Dirac equation in the Kerr-Newman background and to show that $J$ is a symmetry operator. In Section 5 we restrict our attention to the free Dirac operator in the Minkowski metric in oblate spheroidal coordinates. In this framework we bring the Dirac equation into the form of a Schr\"{o}dinger equation and we construct explicitly a positive scalar product respect to which the Hamilton operator is hermitian. In Section 6 the main results are Theorem 6.2, where we prove the completeness of Chandrasekhar ansatz and Theorem 6.3, which gives the integral representation for the Dirac propagator in the Minkowski metric in oblate spheroidal coordinates.   

\section{Preliminaries} \label{section1}
In the Newman-Penrose formalism \cite{New} in two-component spinor notation the Dirac equation coupled to a general gravitational field and a $4$-vector field $\textbf{V}$ is given in Planck units $\hbar=c=G=1$ by \cite{Page}
\begin{eqnarray}
(\nabla_{AA^{'}}+ieV_{AA^{'}})P^{A}+i\mu_{*}Q^{*}_{A^{'}}&=&0 \label{unopage}\\
(\nabla_{AA^{'}}-ieV_{AA^{'}})Q^{A}+i\mu_{*}P^{*}_{A^{'}}&=&0, \label{duepage}
\end{eqnarray}
where $\nabla_{AA^{'}}$ is the symbol for covariant differentiation, $e$ is the charge or coupling constant of the Dirac particle to the vector field $\textbf{V}$, $\mu_{*}$ is the particle mass $m_{e}$ divided by $\sqrt{2}$ and $P^{A}$ and $Q^{A}$ are the two-component spinors representing the wave function. Here, the asterisk used as a superscript denotes complex conjugation. Notice that the factor $2^{-\frac{1}{2}}$ in the definition of the mass $\mu_{*}$ is due to the fact, that the Pauli matrices as defined in the Newman-Penrose formalism differ from their usual definitions by the factor $\sqrt{2}$. Moreover, the vector potential enters with opposite signs in (\ref{unopage}) and (\ref{duepage}) in order that gauge invariance be preserved, since the spinors in the above equations are related by complex conjugation.\\
According to \cite{New}, we denote with $\xi_{a}^{A}$ a basis for the spinor space and with ${\xi_{a^{'}}^{*}}^{A^{'}}$ a basis for the conjugate spinor space. To the spinor basis we can associate at each point of the space-time a null tetrad $(\mathbf{l},\mathbf{n},\mathbf{m},\mathbf{m}^{*})$ obeying the orthogonality relations $\mathbf{l}\cdot\mathbf{n}=1$, $\mathbf{m}\cdot\mathbf{m^{*}}=-1$ and $\mathbf{l}\cdot\mathbf{m}=\mathbf{l}\cdot\mathbf{m^{*}}=\mathbf{n}\cdot\mathbf{m}=\mathbf{n}\cdot\mathbf{m^{*}}=0$. Furthermore, the covariant derivative of a spinor $\xi^{A}$ can be expressed in terms of its components along the spinor basis $\xi_{a}^{A}$ as follows \cite{New}, \cite{Roger}
\begin{equation*}
\xi_{a}^{A}\xi_{a^{'}}^{A^{'}}\xi_{b}^{B}\nabla_{AA^{'}}\xi^{B}=\partial_{aa^{'}}\xi^{b}+\Gamma^{b}_{caa^{'}}\xi^{c},
\end{equation*}
where $\Gamma^{b}_{caa^{'}}$ are the spin coefficients and 
\begin{equation} \label{dirder1}
\partial_{00^{'}}:=D=l^{b}\frac{\partial}{\partial x^{b}},\quad \partial_{11^{'}}:=\widetilde{\Delta}:=n^{b}\frac{\partial}{\partial x^{b}}
\end{equation}
\begin{equation} \label{dirder2}
\partial_{01^{'}}:=\delta=m^{b}\frac{\partial}{\partial x^{b}},\quad \partial_{10^{'}}:=\delta^{*}:={m^{*}}^{b}\frac{\partial}{\partial x^{b}}
\end{equation}
are the directional derivatives along $\mathbf{l}$, $\mathbf{n}$, $\mathbf{m}$, and $\mathbf{m}^{*}$. Following \cite{Page} and letting
\begin{equation*}
P^{0}=:F_{1},\quad P^{1}=:F_{2},\quad {Q^{*}}^{1^{'}}=:G_{1},\quad {Q^{*}}^{0^{'}}=:-G_{2},
\end{equation*}
equations (\ref{unopage}) and (\ref{duepage}) can be brought into the form
 \begin{equation} \label{uno}
\mathcal{O}_{D}\Psi=0
\end{equation}
 with
\begin{equation} \label{operator}
\hspace{-2.5cm}\mathcal{O}_{D}=\left(\begin{array}{cccc} 
\scriptstyle -im_{e}&\scriptstyle 0&\scriptstyle \widetilde{\Delta}+\mu^{*}-\gamma^{*}+ieV_{b}n^{b}& \scriptstyle -(\delta^{*}+\beta^{*}-\tau^{*}+ieV_{b}m^{b})\\
\scriptstyle 0&\scriptstyle -im_{e}&\scriptstyle -(\delta+\pi^{*}-\alpha^{*}+ieV_{b}m^{b})&\scriptstyle D+\epsilon^{*}-\rho^{*}+ieV_{b}l^{b}\\
\scriptstyle D+\epsilon-\rho+ieV_{b}l^{b}&\scriptstyle \delta^{*}+\pi-\alpha+ieV_{b}{m^{*}}^{b}&\scriptstyle -im_{e}&\scriptstyle 0\\
\scriptstyle \delta+\beta-\tau+ieV_{b}{{m}^{*}}^{b}&\scriptstyle \widetilde{\Delta}+\mu-\gamma+ieV_{b}n^{b}&\scriptstyle 0&\scriptstyle -im_{e}
\end{array}\right)
\end{equation}
and $\Psi=(F_{1},F_{2},G_{1},G_{2})^{T}$. In what follows, we consider the Dirac equation in the Kerr-Newman metric, i.e. in the presence of a rotating charged black hole. In Boyer-Lindquist coordinates $(t,r,\vartheta,\varphi)$ with $r>0$, $0\leq\vartheta\leq\pi$, $0\leq\varphi<2\pi$ the Kerr-Newman metric is given by \cite{Cart}
\begin{equation} \label{KN}
\hspace{-1cm}ds^{2}=\frac{\Delta}{\Sigma}(dt-a\sin^{2}\vartheta d\varphi)^{2}-\Sigma\left(\frac{dr^2}{\Delta}+d\vartheta^2\right)-\frac{\sin^{2}\vartheta}{\Sigma}[adt-(r^2+a^2)d\varphi]^2
\end{equation}
with
\begin{equation*}
\Sigma:=\Sigma(r,\theta)=r^2+a^2\cos^{2}\theta, \quad \Delta:=\Delta(r)=r^2-2Mr+a^2+Q^2,
\end{equation*}
where $M$, $a$ and $Q$ are the mass, the angular momentum per unit mass and the charge of the black hole, respectively. In the non-extreme case $M^2>a^2+Q^2$ the function $\Delta$ has two distinct zeros, namely,
\begin{equation*} 
r_{0}=M-\sqrt{M^2-a^2-Q^2} \quad\mbox{and} \quad r_{1}=M+\sqrt{M^2-a^2-Q^2},
\end{equation*}
the first one corresponding to the Cauchy horizon and the second to the event horizon, while in the extreme case $M^2=a^2+Q^2$ Cauchy horizon and event horizon coincide, since $\Delta$ has a double root at $r_{1}^{*}=M$. Throughout our work we will consider the case $M^2> a^2+Q^2$ and restrict our attention to the region $r>r_1$ outside the event horizon. Hence, $\Delta$ will be always positive. Notice that by setting $M=Q=0$ in (\ref{KN}) the Kerr-Newman metric goes over into the Minkowski metric in oblate spheroidal coordinates (OSC), namely
\begin{equation} \label{Mink}
\hspace{-1cm}ds^{2}=dt^{2}-\frac{\Sigma}{\hat{\Delta}}dr^{2}-\hat{\Delta}\left(\frac{\Sigma}{\hat{\Delta}}d\vartheta^{2}+\sin^{2}\vartheta d\varphi^{2}\right),
\end{equation}
with $\hat{\Delta}=r^2+a^2$. In fact, by means of the coordinate transformation
\begin{equation} \label{OSCco}
x=\sqrt{r^2+a^2}\sin{\vartheta}\cos{\varphi},\quad y=\sqrt{r^2+a^2}\sin{\vartheta}\sin{\varphi},\quad z=r\cos{\vartheta}
\end{equation}
(\ref{Mink}) can be reduced to the Minkowski metric in Cartesian coordinates. Moreover, in the OSC the surfaces $r=$const are confocal ellipsoids, while the surfaces $\vartheta=$const are represented by hyperbolids of one sheet, confocal to the ellipsoids. Since sgn$(z)=$sgn$(cos{\vartheta})$ with $\vartheta\in[0,\pi]$, the surface $\vartheta=$const is actually given by a half-hyperboloid, truncated at its waist and laying in the half-space $z\lessgtr 0$ according as $\vartheta\lessgtr\frac{\pi}{2}$. At this point it is interesting to observe that at $r=0$ the ellipsoids degenerate to the disk $x^2+y^2=a^2\sin^{2}{\vartheta}$, whose boundary $\{(r,\vartheta,\varphi)\vert r=0,\vartheta=\frac{\pi}{2}\}$ (and hence $\Sigma=0$) corresponds to the set of points at which the Kerr-Newman metric becomes truly singular. Furthermore, the surfaces $\varphi=$const look like bent planes, which are approximately vertical for large $r$, but flatten out and become horizontal at the edge of the disk. For further details we refer to \cite{Flamm} and \cite{Boy}.\\
In order to write out explicitly (\ref{uno}) for the Kerr-Newman metric in Boyer-Lindquist coordinates, we make use of the Kinnersley tetrad \cite{Kinn}, namely 
\begin{eqnarray} 
l^{b}&=&\left(\frac{r^2+a^2}{\Delta},1,0,\frac{a}{\Delta}\right)\label{t1}\\
n^{b}&=&\frac{1}{2\Sigma}\left(r^2+a^2,-\Delta,0,a\right)\label{t2}\\ m^{b}&=&\frac{1}{\sqrt{2}(r+ia\cos{\vartheta})}\left(ia\sin{\vartheta},0,1,i\csc{\vartheta}\right). \label{t3}
\end{eqnarray}
Then, the non vanishing spin coefficients in (\ref{operator}) are \cite{Bose} 
\begin{equation*}
\rho=-\frac{1}{r-ia\cos{\vartheta}},\quad \beta=\frac{\rho^{*}\cot{\vartheta}}{2\sqrt{2}},\quad \hspace{2mm}\mu=\frac{\Delta\rho}{2\Sigma}
\end{equation*}
\begin{equation*}
\pi=\frac{ia}{\sqrt{2}}\rho^2\sin{\vartheta},\quad \hspace{5mm}\tau=-\frac{ia\sin{\vartheta}}{\sqrt{2}\Sigma},\quad \gamma=\mu+\frac{r-M}{2\Sigma}
\end{equation*}
and
\begin{equation*}
\alpha=\pi-\beta^{*}=\frac{\rho}{\sqrt{2}}\left(ia\rho\sin{\vartheta}+\frac{1}{2}\cot{\vartheta}\right),
\end{equation*}
where $\rho^{*}$ denotes the complex conjugate of $\rho$. Notice that while the vanishing of $\epsilon$ is due to the particular choice of the tetrad (\ref{t1}), (\ref{t2}), (\ref{t3}), the other spin coefficients $\kappa$, $\sigma$, $\lambda$ and $\nu$ are zero according to the Goldberg-Sachs theorem \cite{Goldberg}. By means of (\ref{dirder1}) and (\ref{dirder2}) the directional derivatives along the tetrad are computed to be 
\begin{eqnarray*}
D&=&D_{0},\quad \hspace{0.9cm}D_{0}:=\frac{\partial}{\partial r}+\frac{1}{\Delta}\left[(r^2+a^2)\frac{\partial}{\partial t}+a\frac{\partial}{\partial\varphi}\right]\\
\widetilde{\Delta}&=&-\frac{\Delta \hat{D}_{0}}{2\Sigma},\quad \hat{D}_{0}:=\frac{\partial}{\partial r}-\frac{1}{\Delta}\left[(r^2+a^2)\frac{\partial}{\partial t}+a\frac{\partial}{\partial\varphi}\right]\\
\delta&=&-\frac{\rho^{*}\hat{\mathcal{L}}_{0}}{\sqrt{2}},\quad \hspace{0.5mm}\mathcal{L}_{0}:=\frac{\partial}{\partial \vartheta}-i\left(a\sin{\vartheta}\frac{\partial}{\partial t}+\csc{\vartheta}\frac{\partial}{\partial\varphi}\right)\\
\delta^{*}&=&-\frac{\rho\mathcal{L}_{0}}{\sqrt{2}},\quad\hspace{2mm} \hat{\mathcal{L}}_{0}:=\frac{\partial}{\partial \vartheta}+i\left(a\sin{\vartheta}\frac{\partial}{\partial t}+\csc{\vartheta}\frac{\partial}{\partial\varphi}\right).
\end{eqnarray*}
From \cite{Mis} we know that the vector potential for a rotating charged black hole is
\begin{equation*}
V_{b}=\frac{1}{\Sigma}(-Qr,0,0,aQr\sin^{2}{\vartheta})
\end{equation*}
and therefore 
\begin{equation*}
V_{b}l^{b}=-\frac{rQ}{\Delta},\quad V_{b}n^{b}=-\frac{rQ}{2\Sigma},\quad V_{b}m^{b}=V_{b}{m^{*}}^{b}=0.
\end{equation*}
Notice that the vector potential $\textbf{V}$ is stationary and axisymmetric. Regarding the elements of the matrix operator $\mathcal{O}_{D}$, we find that
\begin{eqnarray*}
\widetilde{\Delta}+\mu^{*}-\gamma^{*}+ie V_{b}n^{b}&=&-\frac{\Delta}{\sqrt{2}\Sigma}\left(\hat{D}_{0}+\frac{r-M}{\Delta}+i\frac{eQr}{\Delta}\right)\\
-(\delta^{*}+\beta^{*}-\tau^{*}+ie V_{b}m^{b})&=&\rho\left(\mathcal{L}_{0}+\frac{1}{2}\cot{\vartheta}+ia\rho^{*}\sin{\vartheta}\right)\\
-(\delta+\pi^{*}-\alpha^{*}+ie V_{b}m^{b})&=&\rho^{*}\left(\hat{\mathcal{L}}_{0}+\frac{1}{2}\cot{\vartheta}\right)\\
D+\epsilon^{*}-\rho^{*}+ie V_{b}l^{b}&=&\sqrt{2}\left(D_{0}-\rho^{*}-i\frac{eQr}{\Delta}\right)\\
D+\epsilon-\rho+ie V_{b}l^{b}&=&\sqrt{2}\left(D_{0}-\rho-i\frac{eQr}{\Delta}\right)\\
\delta^{*}+\pi-\alpha+ie V_{b}{m^{*}}^{b}&=&-\rho\left(\mathcal{L}_{0}+\frac{1}{2}\cot{\vartheta}\right)\\
\delta+\beta-\tau+ie V_{b}{m^{*}}^{b}&=&-\rho^{*}\left(\hat{\mathcal{L}}_{0}+\frac{1}{2}\cot{\vartheta}-ia\rho\sin{\vartheta}\right)\\
\widetilde{\Delta}+\mu-\gamma+ie V_{b}n^{b}&=&-\frac{\Delta}{\sqrt{2}\Sigma}\left(\hat{D}_{0}+\frac{r-M}{\Delta}+i\frac{eQr}{\Delta}\right).
\end{eqnarray*}
According to \cite{Cart1} we may replace the Dirac equation $\mathcal{O}_{D}\Psi(t,r,\vartheta,\varphi)=0$ by a modified but equivalent equation
\begin{equation} \label{vdoppio}
\mathcal{W}\hat{\psi}(t,r,\vartheta,\varphi)=0\quad\mbox{with}\quad \mathcal{W}=\Gamma S^{-1}\mathcal{O}_{D}S,
\end{equation}
where $\hat{\psi}=S^{-1}\Psi=(\hat{F}_{1},\hat{F}_{2},\hat{G}_{1},\hat{G}_{2})^{T}$ and $\Gamma$ and $S$ are non singular $4\times 4$ matrices, whose elements may depend on the variables $r$ and $\vartheta$.

\section{The decomposition of the operator $\mathcal{W}$}
In this section we show that it is possible to find non singular matrices $S$ and $\Gamma$ such that the operator $\mathcal{W}$ decomposes into the sum of an operator containing only derivatives respect to the variables $t$, $r$ and $\varphi$ and of an operator involving only derivatives respect to $t$, $\vartheta$ and $\varphi$. There are several choices of $S$ and $\Gamma$, which accomplish this. Here, we look for matrices $S$ and $\Gamma$ such that they simplify later on the transformation of equation (\ref{vdoppio}) into its Schr\"{o}dinger form.\\
\newtheorem{theorem}{Theorem}[section]
     \newtheorem{lemma}[theorem]{Lemma}
     \newenvironment{Proof.}[1][Proof.]{\begin{trivlist}
     \item[\hskip \labelsep {\bfseries #1}]}{\end{trivlist}}
      
\begin{lemma} \label{lemma1}
Let $r$ be positive with $r>r_{1}$. There exist a non singular $4\times 4$ matrix 
\begin{equation*}
S=c\hspace{0.5mm}\mbox{diag}(-\rho,2^{\frac{1}{2}}\Delta^{-\frac{1}{2}},2^{\frac{1}{2}}\Delta^{-\frac{1}{2}},-\rho^{*}),\quad\mbox{det}(S)=\frac{c^4}{\Sigma\Delta}
\end{equation*} 
with $0\neq c\in\mathbb{C}$ and a non singular $4\times 4$ matrix
\begin{equation*}
\Gamma=\mbox{diag}({\rho^{*}}^{-1},-{\rho^{*}}^{-1},-\rho^{-1},\rho^{-1}),\quad\hspace{0.6cm}\mbox{det}(\Gamma)=\Sigma
\end{equation*}
such that
\begin{equation} \label{unoo}
\mathcal{W}=\mathcal{W}_{(t,r,\varphi)}+\mathcal{W}_{(t,\vartheta,\varphi)}
\end{equation}
with
\begin{equation} \label{due}
\mathcal{W}_{(t,r,\varphi)}=\left( \begin{array}{cccc}
                            im_{e}r&0&\sqrt{\Delta}\mathcal{D}_{+}&0\\
                            0&-im_{e}r&0&\sqrt{\Delta}\mathcal{D}_{-}\\
                            \sqrt{\Delta}\mathcal{D}_{-}&0&-im_{e}r&0\\
                             0&\sqrt{\Delta}\mathcal{D}_{+}&0&im_{e}r
                            \end{array} \right)
\end{equation}
\begin{equation} \label{3}
\mathcal{W}_{(t,\vartheta,\varphi)}=\left( \begin{array}{cccc}
                            -am_{e}\cos{\vartheta}&0&0&\mathcal{L}_{+}\\
                            0&am_{e}\cos{\vartheta}&-\mathcal{L}_{-}&0\\
                            0&\mathcal{L}_{+}&-am_{e}\cos{\vartheta}&0\\
                            -\mathcal{L}_{-}&0&0&am_{e}\cos{\vartheta}
                            \end{array} \right)
\end{equation}
where
\begin{eqnarray*}
\mathcal{D}_{\pm}:&=&\frac{\partial}{\partial r}\mp\frac{1}{\Delta}\left[(r^2+a^2)\frac{\partial}{\partial t}+a\frac{\partial}{\partial\varphi}-ieQr\right]\\
\mathcal{L}_{\pm}:&=&\frac{\partial}{\partial \vartheta}+\frac{1}{2}\cot{\vartheta}\mp i\left(a\sin{\vartheta}\frac{\partial}{\partial t}+\csc{\vartheta}\frac{\partial}{\partial\varphi}\right).
\end{eqnarray*}
\end{lemma}
\begin{Proof.}
Let us define the matrix
\begin{equation*}
S:=\mbox{diag}(h(r,\vartheta),\Lambda(r,\vartheta),\sigma(r,\vartheta),\gamma(r,\vartheta))
\end{equation*}
such that det$(S)\neq 0$ and $h$, $\Lambda$, $\sigma$, $\gamma$ at least $\mathcal{C}^{1}((r_{1},+\infty)\times[0,\pi])$. The equation $\mathcal{O}_{D}S\hat{\psi}=0$ gives rise to the system
\begin{eqnarray}
\hspace{-2.5cm}-im_{e}h\hat{F}_{1}+(\widetilde{\Delta}+\mu^{*}-\gamma^{*}+ie V_{b}n^{b})\sigma\hat{G}_{1}-(\delta^{*}+\beta^{*}-\tau^{*}+ie V_{b}m^{b})\gamma\hat{G}_{2}&=&0 \label{A}\\
\hspace{-2.5cm}-im_{e}\Lambda\hat{F}_{2}-(\delta+\pi^{*}-\alpha^{*}+ie V_{b}m^{b})\sigma\hat{G}_{1}+(D+\epsilon^{*}-\rho^{*}+ie V_{b}l^{b})\gamma\hat{G}_{2}&=&0 \label{B}\\
\hspace{-2.5cm}-im_{e}\sigma\hat{G}_{1}+(D+\epsilon-\rho+ie V_{b}l^{b})h\hat{F}_{1}+(\delta^{*}+\pi-\alpha+ie V_{b}{m^{*}}^{b})\Lambda\hat{F}_{2}&=&0 \label{C}\\
\hspace{-2.5cm}-im_{e}\gamma\hat{G}_{2}+(\delta+\beta-\tau+ie V_{b}{m^{*}}^{b})h\hat{F}_{1}+(\widetilde{\Delta}+\mu-\gamma+ie V_{b}n^{b})\Lambda\hat{F}_{2}&=&0 \label{D}.
\end{eqnarray}
Starting with (\ref{A}), we find that
\begin{eqnarray}
\hspace{-2cm}(\widetilde{\Delta}+\mu^{*}-\gamma^{*}+ie V_{b}n^{b})\sigma\hat{G}_{1}&=&-\frac{\Delta}{\sqrt{2}\Sigma}\left(\sigma \mathcal{D}_{+}+\frac{\partial\sigma}{\partial r}+\frac{r-M}{\Delta}\sigma\right)\hat{G}_{1} \label{A1}\\
\hspace{-2cm}-(\delta^{*}+\beta^{*}-\tau^{*}+ie V_{b}m^{b})\gamma\hat{G}_{2}&=&\rho\left(\gamma\mathcal{L}_{+}+\frac{\partial\gamma}{\partial \vartheta}+ia\rho^{*}\sin{\vartheta}\hspace{1mm}\gamma\right)\hat{G}_{2} \label{A2}
\end{eqnarray}
with
\begin{eqnarray*}
\mathcal{D}_{+}:&=&\frac{\partial}{\partial r}-\frac{1}{\Delta}\left[(r^2+a^2)\frac{\partial}{\partial t}+a\frac{\partial}{\partial\varphi}-ieQr\right]\\
\mathcal{L}_{+}:&=&\frac{\partial}{\partial \vartheta}+\frac{1}{2}\cot{\vartheta}-i\left(a\sin{\vartheta}\frac{\partial}{\partial t}+\csc{\vartheta}\frac{\partial}{\partial\varphi}\right).
\end{eqnarray*}
In order to obtain $\mathcal{W}$ as in the statement of this Lemma, we impose in (\ref{A1}) that
\begin{equation*}
\frac{\partial\sigma}{\partial r}+\frac{r-M}{\Delta}\sigma=0,
\end{equation*}
whose solution is given by $\sigma(r,\vartheta)=\widetilde{\sigma}(\vartheta)\Delta^{-\frac{1}{2}}$ with $\widetilde{\sigma}(\vartheta)\neq 0$ for every $\vartheta\in[0,\pi]$, while in (\ref{A2}) we require that
\begin{equation*}
\frac{\partial\gamma}{\partial \vartheta}+ia\rho^{*}\sin{\vartheta}\hspace{1mm}\gamma=0,
\end{equation*} 
whose solution is $\gamma(r,\vartheta)=-\widetilde{\gamma}(r)\rho^{*}$ with $\widetilde{\gamma}(r)\neq 0$ for every $r\in(r_{1},+\infty)$. Hence, (\ref{A1}) and (\ref{A2}) simplify to
\begin{eqnarray}
(\widetilde{\Delta}+\mu^{*}-\gamma^{*}+ie V_{b}n^{b})\sigma\hat{G}_{1}&=&-\frac{\sqrt{2}\Delta^{\frac{1}{2}}\widetilde{\sigma}(r)}{2\Sigma}\mathcal{D}_{+} \hat{G}_{1} \label{AA1}\\
-(\delta^{*}+\beta^{*}-\tau^{*}+ie V_{b}m^{b})\gamma\hat{G}_{2}&=&-\frac{\widetilde{\gamma}(r)}{\Sigma}\mathcal{L}_{+}\hat{G}_{2}. \label{AA2}
\end{eqnarray}
Regarding (\ref{B}), we have
\begin{eqnarray}
-(\delta+\pi^{*}-\alpha^{*}+ie V_{b}m^{b})\sigma\hat{G}_{1}&=&\frac{\rho^{*}}{\Delta^{\frac{1}{2}}}\left(\widetilde{\sigma}\mathcal{L}_{-}+\frac{d\widetilde{\sigma}}{d\vartheta}\right)\hat{G}_{1} \label{A3}\\
(D+\epsilon^{*}-\rho^{*}+ie V_{b}l^{b})\gamma\hat{G}_{2}&=&\sqrt{2}\left(\gamma \mathcal{D}_{-}+\frac{\partial\gamma}{\partial r}-\rho^{*}\gamma\right)\hat{G}_{2}, \label{A4}
\end{eqnarray}
where
\begin{eqnarray*}
\mathcal{D}_{-}:&=&\frac{\partial}{\partial r}+\frac{1}{\Delta}\left[(r^2+a^2)\frac{\partial}{\partial t}+a\frac{\partial}{\partial\varphi}-ieQr\right]\\
\mathcal{L}_{-}:&=&\frac{\partial}{\partial \vartheta}+\frac{1}{2}\cot{\vartheta}+i\left(a\sin{\vartheta}\frac{\partial}{\partial t}+\csc{\vartheta}\frac{\partial}{\partial\varphi}\right).
\end{eqnarray*}
By imposing $\widetilde{\sigma}(\vartheta)=c_{3}$ with some constant $0\neq c_{3}\in\mathbb{C}$ and observing that $\gamma(r,\vartheta)=-c_{4}\rho^{*}$ with $0\neq c_{4}\in\mathbb{C}$ sets the last two terms in the parenthesis on the r.h.s. of (\ref{A4}) equal to zero, (\ref{AA1}), (\ref{AA2}), (\ref{A3}) and (\ref{A4}) become
\begin{eqnarray*}
(\widetilde{\Delta}+\mu^{*}-\gamma^{*}+ie V_{b}n^{b})\sigma\hat{G}_{1}&=&-\frac{c_{3}\sqrt{2}\Delta^{\frac{1}{2}}}{2\Sigma}\mathcal{D}_{+}\hat{G}_{1}\\
-(\delta^{*}+\beta^{*}-\tau^{*}+ie V_{b}m^{b})\gamma\hat{G}_{2}&=&-\frac{c_{4}}{\Sigma}\mathcal{L}_{+}\hat{G}_{2}\\
-(\delta+\pi^{*}-\alpha^{*}+ie V_{b}m^{b})\sigma\hat{G}_{1}&=&\frac{c_{3}\rho^{*}}{\Delta^{\frac{1}{2}}}\mathcal{L}_{-}\hat{G}_{1}\\
(D+\epsilon^{*}-\rho^{*}+ie V_{b}l^{b})\gamma\hat{G}_{2}&=&-c_{4}\sqrt{2}\rho^{*}\mathcal{D}_{-}\hat{G}_{2}.
\end{eqnarray*} 
Concerning (\ref{C}), a short computation gives
\begin{eqnarray}
(D+\epsilon-\rho+ie V_{b}l^{b})h\hat{F}_{1}&=&\sqrt{2}\left(h\mathcal{D}_{-}+\frac{\partial h}{\partial r}-\rho h\right)\hat{F}_{1} \label{A5}\\
(\delta^{*}+\pi-\alpha+ie V_{b}{m^{*}}^{b})\Lambda\hat{F}_{2}&=&-\rho\left(\Lambda\mathcal{L}_{+}+\frac{\partial\Lambda}{\partial\vartheta}\right)\hat{F}_{2} \label{A6}
\end{eqnarray}
and by similar reasonings as we did for (\ref{A3}) and (\ref{A4}), we obtain $h(r,\vartheta)=-\rho\widetilde{h}(\vartheta)$ and $\Lambda(r,\vartheta)=c_{2}\widetilde{\lambda}(r)$ with $0\neq c_{2}\in\mathbb{C}$. Finally, regarding (\ref{D}), we get
\begin{eqnarray}
\hspace{-5mm}(\delta+\beta-\tau+ie V_{b}{m^{*}}^{b})h\hat{F}_{1}&=&-\rho^{*}\left(h\mathcal{L}_{-}+\frac{\partial h}{\partial\vartheta}-ia\rho\sin{\vartheta}\hspace{0.5mm}h\right)\hat{F}_{1} \label{A7} \\
\hspace{-5mm}(\widetilde{\Delta}+\mu-\gamma+ie V_{b}n^{b})\Lambda\hat{F}_{2}&=&-\frac{\sqrt{2}\Delta}{2\Sigma}\left(\Lambda \mathcal{D}_{+}+\frac{\partial\Lambda}{\partial r}+\frac{r-M}{\Delta}\Lambda\right)\hat{F}_{2} \label{A8}.
\end{eqnarray}
By proceeding similarly as we did for (\ref{A3}) and (\ref{A4}), we find that $\Lambda(r,\vartheta)=c_{2}\Delta^{-\frac{1}{2}}$ and $h(r,\vartheta)=-c_{1}\rho$ with $0\neq c_{1}\in\mathbb{C}$. Hence, (\ref{A5}), (\ref{A6}), (\ref{A7}) and (\ref{A8}) become
\begin{eqnarray*}
(D+\epsilon-\rho+ie V_{b}l^{b})h\hat{F}_{1}&=&-c_{1}\sqrt{2}\rho \mathcal{D}_{-}\hat{F}_{1}\\
(\delta^{*}+\pi-\alpha+ie V_{b}{m^{*}}^{b})\Lambda\hat{F}_{2}&=&-c_{2}\rho\Delta^{-\frac{1}{2}}\mathcal{L}_{+}\hat{F}_{2}\\
(\delta+\beta-\tau+ie V_{b}{m^{*}}^{b})h\hat{F}_{1}&=&\frac{c_{1}}{\Sigma}\mathcal{L}_{-}\hat{F}_{1}\\
(\widetilde{\Delta}+\mu-\gamma+ie V_{b}n^{b})\Lambda\hat{F}_{2}&=&-\frac{c_{2}\sqrt{2}\Delta^{\frac{1}{2}}}{2\Sigma}\mathcal{D}_{+}\hat{F}_{2}.
\end{eqnarray*}
Then, the operator $S^{-1}\mathcal{O}S$ is computed to be
\begin{equation*} 
\hspace{-1cm}S^{-1}\mathcal{O}S=\left( \begin{array}{cccc}
\scriptstyle -im_{e}&\scriptstyle 0&\frac{\sqrt{2}c_{3}\rho^{*}\Delta^{\frac{1}{2}}}{2c_{1}}\scriptstyle \mathcal{D}_{+}&\frac{c_{4}\rho^{*}}{c_{1}}\scriptstyle \mathcal{L}_{+}\\
\scriptstyle 0&\scriptstyle -im_{e}&\frac{c_{3}\rho^{*}}{c_{2}}\scriptstyle \mathcal{L}_{-}&-\frac{\sqrt{2}c_{4}\rho^{*}\Delta^{\frac{1}{2}}}{c_{2}}\scriptstyle \mathcal{D}_{-}\\
-\frac{\sqrt{2}c_{1}\rho\Delta^{\frac{1}{2}}}{c_{3}}\scriptstyle \mathcal{D}_{-}&-\frac{c_{2}\rho}{c_{3}}\scriptstyle \mathcal{L}_{+}&\scriptstyle -im_{e}&\scriptstyle 0\\
-\frac{c_{1}\rho}{c_{4}}\scriptstyle \mathcal{L}_{-}&\frac{\sqrt{2}c_{2}\rho\Delta^{\frac{1}{2}}}{2c_{4}}\scriptstyle \mathcal{D}_{+}&\scriptstyle 0&\scriptstyle -im_{e}
                           \end{array} \right).
\end{equation*}
By choosing a non singular $4\times 4$ matrix $\Gamma$ as follows
\begin{equation*}
\Gamma=\mbox{diag}({\rho^{*}}^{-1},-{\rho^{*}}^{-1},-\rho^{-1},\rho^{-1}),
\end{equation*}
it can be checked that
\begin{equation*} 
\hspace{-2cm}\mathcal{W}=\Gamma S^{-1}\mathcal{O}S=\left( \begin{array}{cccc}
\scriptstyle im_{e}r-am_{e}\cos{\vartheta}&\scriptstyle 0&\frac{\sqrt{2}c_{3}}{2c_{1}}\scriptstyle \sqrt{\Delta}\mathcal{D}_{+}&\frac{c_{4}}{c_{1}}\scriptstyle \mathcal{L}_{+}\\
\scriptstyle 0&\scriptstyle -im_{e}r+am_{e}\cos{\vartheta}&-\frac{c_{3}}{c_{2}}\scriptstyle \mathcal{L}_{-}&-\frac{\sqrt{2}c_{4}}{c_{2}}\scriptstyle \sqrt{\Delta}\mathcal{D}_{-}\\
\frac{\sqrt{2}c_{1}}{c_{3}}\scriptstyle \sqrt{\Delta}\mathcal{D}_{-}&\frac{c_{2}}{c_{3}}\scriptstyle \mathcal{L}_{+}&\scriptstyle -im_{e}r-am_{e}\cos{\vartheta}&\scriptstyle 0\\
-\frac{c_{1}}{c_{4}}\scriptstyle \mathcal{L}_{-}&\frac{\sqrt{2}c_{2}}{2c_{4}}\scriptstyle \sqrt{\Delta}\mathcal{D}_{+}&\scriptstyle 0&\scriptstyle im_{e}r+am_{e}\cos{\vartheta}
                           \end{array} \right).
\end{equation*}
If we impose the following conditions
\begin{equation*}
c_{4}=\frac{c_{3}}{\sqrt{2}}=\frac{c_{2}}{\sqrt{2}}=c_{1}=:c,
\end{equation*}
we finally have
\begin{equation*} 
\mathcal{W}=\left( \begin{array}{cccc}
\scriptstyle im_{e}r-am_{e}\cos{\vartheta}&\scriptstyle 0&\scriptstyle \sqrt{\Delta}\mathcal{D}_{+}&\scriptstyle \mathcal{L}_{+}\\
\scriptstyle 0&\scriptstyle -im_{e}r+am_{e}\cos{\vartheta}&-\scriptstyle \mathcal{L}_{-}&\scriptstyle \sqrt{\Delta}\mathcal{D}_{-}\\
\scriptstyle \sqrt{\Delta}\mathcal{D}_{-}&\scriptstyle \mathcal{L}_{+}&\scriptstyle -im_{e}r-am_{e}\cos{\vartheta}&\scriptstyle 0\\
-\scriptstyle \mathcal{L}_{-}&\scriptstyle \sqrt{\Delta}\mathcal{D}_{+}&\scriptstyle 0&\scriptstyle im_{e}r+am_{e}\cos{\vartheta}
                           \end{array} \right).
\end{equation*}
This completes the proof.\\
\vspace{-0.9cm}
\begin{equation*}
\hspace{12cm}\square
\end{equation*}
\end{Proof.}
Clearly, the result obtained in Lemma~\ref{lemma1} is also valid in the extreme case, where $\Delta$ is simply $(r-M)^2$. Without loss of generality we set $c=1$. The next result, which holds for the Dirac equation in OSC, follows directly from Lemma~\ref{lemma1} by setting $M=Q=0$.

     \newtheorem{proposition}[theorem]{Proposition}
     \newtheorem{corollary}[theorem]{Corollary}

\begin{corollary} \label{corollary1}
Let $r>0$. There exist a non singular $4\times 4$ matrix 
\begin{equation*}
S^{(OSC)}=\hspace{0.5mm}\mbox{diag}(-\rho,2^{\frac{1}{2}}\hat{\Delta}^{-\frac{1}{2}},2^{\frac{1}{2}}\hat{\Delta}^{-\frac{1}{2}},-\rho^{*}),\quad\mbox{det}(S^{(OSC)})=\frac{1}{\Sigma\hat{\Delta}}
\end{equation*} 
and a non singular $4\times 4$ matrix
\begin{equation*}
\Gamma=\mbox{diag}({\rho^{*}}^{-1},-{\rho^{*}}^{-1},-\rho^{-1},\rho^{-1}),\quad\hspace{0.6cm}\mbox{det}(\Gamma)=\Sigma
\end{equation*}
such that
\begin{equation} \label{unostern}
\mathcal{W}^{(OSC)}=\mathcal{W}_{(t,r,\varphi)}^{(OSC)}+\mathcal{W}_{(t,\vartheta,\varphi)}^{(OSC)}
\end{equation}
with
\begin{equation} \label{duestern}
\hspace{-2cm}\mathcal{W}^{(OSC)}_{(t,r,\varphi)}=\left( \begin{array}{cccc}
                            im_{e}r&0&\sqrt{\hat{\Delta}}\widetilde{\mathcal{D}}_{+}&0\\
                            0&-im_{e}r&0&\sqrt{\hat{\Delta}}\widetilde{\mathcal{D}}_{-}\\
                            \sqrt{\hat{\Delta}}\widetilde{\mathcal{D}}_{-}&0&-im_{e}r&0\\
                             0&\sqrt{\hat{\Delta}}\widetilde{\mathcal{D}}_{+}&0&im_{e}r
                            \end{array} \right),\quad \mathcal{W}^{(OSC)}_{(t,\vartheta,\varphi}=\mathcal{W}_{(t,\vartheta,\varphi)}
\end{equation}
where $\mathcal{W}_{(t,\vartheta,\varphi)}$ is given by (\ref{3}), $\hat{\Delta}=r^2+a^2$ and
\begin{equation*}
\widetilde{\mathcal{D}}_{\pm}:=\frac{\partial}{\partial r}\mp\left(\frac{\partial}{\partial t}+\frac{a}{\hat{\Delta}}\frac{\partial}{\partial\varphi}\right) .
\end{equation*}
\end{corollary}
Notice that the angular part of the operator $\mathcal{W}_{(t,\vartheta,\varphi)}$ remains unchanged, when we treat the Dirac equation in the Minkowski space-time with oblate spheroidal coordinates. Turning back to the Kerr-Newman metric, it can be easily checked that
\begin{equation} \label{4}
\left[\mathcal{W}_{(t,r,\varphi)},\mathcal{W}_{(t,\vartheta,\varphi)}\right]=0,
\end{equation} 
which also implies
\begin{equation*}
\left[\mathcal{W}_{(t,r,\varphi)},\mathcal{W}\right]=\left[\mathcal{W}_{(t,\vartheta,\varphi)},\mathcal{W}\right]=0.
\end{equation*}
An analogous commutator holds for the Dirac equation in OSC. Going back to Lemma~\ref{lemma1}, we observe that the matrix $\Gamma$ splits into the sum
\begin{equation} \label{cinque}
\Gamma=\Gamma_{(r)}+\Gamma_{(\vartheta)},
\end{equation}
where $\Gamma_{(r)}=\mbox{diag}(-r,r,r,-r)$ and $\Gamma_{(\vartheta)}=ia\mbox{diag}(-\cos{\vartheta},\cos{\vartheta},-\cos{\vartheta},\cos{\vartheta})$ satisfy the commutation relations
\begin{equation} \label{sei}
\left[\Gamma_{(r)},\Gamma_{(\vartheta)}\right]=0,
\end{equation}
\begin{equation} \label{sette}
\left[\Gamma_{(r)},\mathcal{W}_{(t,\vartheta,\varphi)}\right]=\left[\Gamma_{(\vartheta)},\mathcal{W}_{(t,r,\varphi)}\right]=0.
\end{equation}

\section{Construction of the symmetry operator  $J$} 
In this section we will make use of Chandrasekhar separation ansatz in order to construct a new operator $J$ for the Dirac equation in the Kerr-Newman metric. We show that $J$ commutes with the Dirac operator $\mathcal{O}_{D}$ given by (\ref{operator}), thus being a symmetry operator for $\mathcal{O}_{D}$. Moreover, the separation constant turns out to be eigenvalue of $J$, whose physical interpretation is that one of the square root of the squared total angular momentum for a Dirac particle in the presence of a charged rotating black hole. \\
Taking into account that the geometry of our problem is axial symmetric and considering stationary waves with energy $\omega$, the $(t,\varphi)$ dependence of the spinors $\hat{\psi}$ entering in (\ref{vdoppio}) is given by
\begin{equation} \label{psi}
\hat{\psi}(t,r,\vartheta,\varphi)=e^{i\omega t}e^{i\left(k+\frac{1}{2}\right)\varphi}\widetilde{\psi}(r,\vartheta)
\end{equation}
where $k\in\mathbb{Z}$ is the azimuthal quantum number of the particle and $\widetilde{\psi}(r,\vartheta)\in\mathbb{C}^{4}$. Inserting (\ref{psi}) in (\ref{vdoppio}), it is straightforward to verify that $\widetilde{\psi}(r,\vartheta)$ satisfies the following equation
\begin{equation} \label{mod}
\left(\mathcal{W}_{(r)}+\mathcal{W}_{(\vartheta)}\right)\widetilde{\psi}=0,
\end{equation}
where
\begin{equation} \label{mod1}
\mathcal{W}_{(r)}=\left( \begin{array}{cccc}
                            im_{e}r&0&\sqrt{\Delta}\hat{\mathcal{D}}_{+}&0\\
                            0&-im_{e}r&0&\sqrt{\Delta}\hat{\mathcal{D}}_{-}\\
                            \sqrt{\Delta}\hat{\mathcal{D}}_{-}&0&-im_{e}r&0\\
                             0&\sqrt{\Delta}\hat{\mathcal{D}}_{+}&0&im_{e}r
                            \end{array} \right)
\end{equation}
and
\begin{equation} \label{mod2}
\mathcal{W}_{(\vartheta)}=\left( \begin{array}{cccc}
                            -am_{e}\cos{\vartheta}&0&0&\hat{\mathcal{L}}_{+}\\
                            0&am_{e}\cos{\vartheta}&-\hat{\mathcal{L}}_{-}&0\\
                            0&\hat{\mathcal{L}}_{+}&-am_{e}\cos{\vartheta}&0\\
                            -\hat{\mathcal{L}}_{-}&0&0&am_{e}\cos{\vartheta}
                            \end{array} \right)
\end{equation}
with
\begin{eqnarray*}
\hat{\mathcal{D}}_{\pm}:&=&\frac{\partial}{\partial r}\mp i\frac{K(r)}{\Delta},\quad \hspace{1.5cm}K(r)=\omega(r^2+a^2)-eQr+\left(k+\frac{1}{2}\right)a\\
\hat{\mathcal{L}}_{\pm}:&=&\frac{\partial}{\partial \vartheta}+\frac{1}{2}\cot{\vartheta}\pm Q(\vartheta),\quad Q(\vartheta)=a\omega\sin{\vartheta}+\left(k+\frac{1}{2}\right)\csc{\vartheta}.
\end{eqnarray*}
By defining $\widetilde{\psi}(r,\vartheta)=(f_{1}(r,\vartheta),f_{2}(r,\vartheta),g_{1}(r,\vartheta),g_{2}(r,\vartheta))^{T}$, equation (\ref{mod}) gives rise to the following systems of first order linear partial differential equations
\begin{eqnarray*}
\sqrt{\Delta}\hat{\mathcal{D}}_{+}g_{1}+im_{e}rf_{1}+(\hat{\mathcal{L}}_{+}g_{2}-am_{e}\cos{\vartheta}f_{1})&=&0\\
\sqrt{\Delta}\hat{\mathcal{D}}_{-}g_{2}-im_{e}rf_{2}-(\hat{\mathcal{L}}_{-}g_{1}-am_{e}\cos{\vartheta}f_{2})&=&0\\
\sqrt{\Delta}\hat{\mathcal{D}}_{-}f_{1}-im_{e}rg_{1}+(\hat{\mathcal{L}}_{+}f_{2}-am_{e}\cos{\vartheta}g_{1})&=&0\\
\sqrt{\Delta}\hat{\mathcal{D}}_{+}f_{2}+im_{e}rg_{2}-(\hat{\mathcal{L}}_{-}f_{1}-am_{e}\cos{\vartheta}g_{2})&=&0.
\end{eqnarray*}
Let us now define
\begin{equation} \label{eins}
f_{1}(r,\vartheta)=\gamma_{1}(r)\delta_{1}(\vartheta),\quad f_{2}(r,\vartheta)=\sigma_{2}(r)\tau_{2}(\vartheta),
\end{equation}
\begin{equation} \label{zwei}
g_{1}(r,\vartheta)=\alpha_{1}(r)\beta_{1}(\vartheta),\quad g_{2}(r,\vartheta)=\epsilon_{2}(r)\mu_{2}(\vartheta).
\end{equation}
If we substitute (\ref{eins}) and (\ref{zwei}) into the above system, it can be easily seen that the requirement of separability implies
\begin{equation*} 
\beta_{1}(\vartheta)=\delta_{1}(\vartheta),\quad \epsilon_{2}(r)=\gamma_{1}(r),\quad\mu_{2}(\vartheta)=\tau_{2}(\vartheta),\quad \alpha_{1}(r)=\sigma_{2}(r).
\end{equation*}
According to the notation used in \cite{Chandra}, we set
\begin{equation*} 
\gamma_{1}(r)=R_{-}(r),\quad \delta_{1}(\vartheta)=S_{-}(\vartheta),\quad \sigma_{2}(r)=R_{+}(r),\quad \tau_{2}(\vartheta)=S_{+}(\vartheta).
\end{equation*}
Hence, we get
\begin{equation*} 
f_{1}(r,\vartheta)=R_{-}(r)S_{-}(\vartheta),\quad f_{2}(r,\vartheta)=R_{+}(r)S_{+}(\vartheta),
\end{equation*}
\begin{equation*}
g_{1}(r,\vartheta)=R_{+}(r)S_{-}(\vartheta),\quad g_{2}(r,\vartheta)=R_{-}(r)S_{+}(\vartheta)
\end{equation*}
and the system of partial differential equations become
\begin{eqnarray}
(\sqrt{\Delta}\hat{\mathcal{D}}_{+}R_{+}+im_{e}rR_{-})S_{-}+(\hat{\mathcal{L}}_{+}S_{+}-am_{e}\cos{\vartheta}S_{-})R_{-}&=&0 \label{c1}\\
(\sqrt{\Delta}\hat{\mathcal{D}}_{-}R_{-}-im_{e}rR_{+})S_{+}-(\hat{\mathcal{L}}_{-}S_{-}-am_{e}\cos{\vartheta}S_{+})R_{+}&=&0 \label{c2}\\
(\sqrt{\Delta}\hat{\mathcal{D}}_{-}R_{-}-im_{e}rR_{+})S_{-}+(\hat{\mathcal{L}}_{+}S_{+}-am_{e}\cos{\vartheta}S_{-})R_{+}&=&0 \label{c3}\\
(\sqrt{\Delta}\hat{\mathcal{D}}_{+}R_{+}+im_{e}rR_{-})S_{+}-(\hat{\mathcal{L}}_{-}S_{-}-am_{e}\cos{\vartheta}S_{+})R_{-}&=&0.\label{c4}
\end{eqnarray}
Introducing four separation constants $\lambda_{1},\cdots,\lambda_{4}$ as follows
\begin{eqnarray}
\hat{\mathcal{L}}_{+}S_{+}-am_{e}\cos{\vartheta}S_{-}&=&-\lambda_{1}S_{-} \label{b1}\\
\hat{\mathcal{L}}_{-}S_{-}-am_{e}\cos{\vartheta}S_{+}&=&+\lambda_{2}S_{+} \label{b2}\\
\hat{\mathcal{L}}_{+}S_{+}-am_{e}\cos{\vartheta}S_{-}&=&-\lambda_{3}S_{-} \label{b3}\\
\hat{\mathcal{L}}_{-}S_{-}-am_{e}\cos{\vartheta}S_{+}&=&+\lambda_{4}S_{+},\label{b4}
\end{eqnarray}
we obtain from the set of equations (\ref{c1}), (\ref{c2}), (\ref{c2}) and (\ref{c4})
\begin{eqnarray}
\sqrt{\Delta}\hat{\mathcal{D}}_{+}R_{+}+im_{e}rR_{-}&=&\lambda_{1}R_{-} \label{d1}\\
\sqrt{\Delta}\hat{\mathcal{D}}_{+}R_{+}+im_{e}rR_{-}&=&\lambda_{4}R_{-} \label{d2}\\
\sqrt{\Delta}\hat{\mathcal{D}}_{-}R_{-}-im_{e}rR_{+}&=&\lambda_{2}R_{+} \label{d3}\\
\sqrt{\Delta}\hat{\mathcal{D}}_{-}R_{-}-im_{e}rR_{+}&=&\lambda_{3}R_{+}.\label{d4}
\end{eqnarray}
Clearly, the systems of equations (\ref{b1}),$\cdots$,(\ref{b4}) and (\ref{d1}),$\cdots$,(\ref{d4}) will be consistent if
\begin{equation*}
\lambda_{1}=\lambda_{2}=\lambda_{3}=\lambda_{4}=:\lambda.
\end{equation*}
Notice that the decoupled equations give rise to the following two systems of linear first order differential equations, namely 
\begin{equation} \label{radial} 
\left( \begin{array}{cc}
     \sqrt{\Delta}\hat{\mathcal{D}}_{-}&-im_{e}r-\lambda\\
     im_{e}r-\lambda&\sqrt{\Delta}\hat{\mathcal{D}}_{+}
           \end{array} \right)\left( \begin{array}{cc}
                                     R_{-} \\
                                     R_{+}
                                     \end{array}\right)=0,
\end{equation}
\begin{equation} \label{angular}
\left( \begin{array}{cc}
     -\hat{\mathcal{L}}_{-} & \lambda+am_{e}\cos\theta\\
                \lambda-am_{e}\cos\theta & \hat{\mathcal{L}}_{+}
           \end{array} \right)\left( \begin{array}{cc}
                                     S_{-} \\
                                     S_{+}
                                     \end{array}\right)=0.
\end{equation} 
Starting from the systems of equations (\ref{b1}),$\cdots$,(\ref{b4}) and (\ref{d1}),$\cdots$,(\ref{d4}), we are now able to construct a new operator $J$. Moreover, we will show that such operator commutes with the Dirac operator $\mathcal{O}_{D}$.\\
By setting $\lambda_{1}=\lambda_{2}=\lambda_{3}=\lambda_{4}=:\lambda$ in (\ref{b1}),$\cdots$,(\ref{b4}) and (\ref{d1}),$\cdots$,(\ref{d4}), multiplying (\ref{b1}) by $-rR_{-}$ and (\ref{d1}) by $ia\cos{\vartheta}S_{-}$, respectively, and summing together the resulting equations, we find that
\begin{equation} \label{h1}
-ia\rho^{*}\cos{\vartheta}\sqrt{\Delta}\hat{\mathcal{D}}_{+}g_{1}+r\rho^{*}\hat{\mathcal{L}}_{+}g_{2}=\lambda f_{1}
\end{equation}
with $\rho^{*}=-(r+ia\cos{\vartheta})^{-1}$. By means of the same method we obtain from the couples of equations ((\ref{b2}),(\ref{d2})) and ((\ref{b4},(\ref{d4}))
\begin{eqnarray}
-ia\rho^{*}\cos{\vartheta}\sqrt{\Delta}\hat{\mathcal{D}}_{-}g_{2}-r\rho^{*}\hat{\mathcal{L}}_{-}g_{1}&=&\lambda f_{2} \label{h2}\\
+ia\rho\cos{\vartheta}\sqrt{\Delta}\hat{\mathcal{D}}_{-}f_{1}+r\rho\hat{\mathcal{L}}_{+}f_{2}&=&\lambda g_{1} \label{h3}\\
+ia\rho\cos{\vartheta}\sqrt{\Delta}\hat{\mathcal{D}}_{+}f_{2}-r\rho\hat{\mathcal{L}}_{-}f_{1}&=&\lambda g_{2} \label{h4}
\end{eqnarray}
with $\rho=-(r-ia\cos{\vartheta})^{-1}$. Equations (\ref{h1}),$\cdots$,(\ref{h4}) give the entries of the matrix operator we are looking for, namely
\begin{equation*} 
\hspace{-2.5cm}\hat{J}=\left( \begin{array}{cccc}
0&0&-ia\rho^{*}\cos{\vartheta}\sqrt{\Delta}\hat{\mathcal{D}}_{+}&r\rho^{*}\hat{\mathcal{L}}_{+}\\
0&0&-r\rho^{*}\hat{\mathcal{L}}_{-}&-ia\rho^{*}\cos{\vartheta}\sqrt{\Delta}\hat{\mathcal{D}}_{-}\\
ia\rho\cos{\vartheta}\sqrt{\Delta}\hat{\mathcal{D}}_{-}&r\rho\hat{\mathcal{L}}_{+}&0&0\\
-r\rho\hat{\mathcal{L}}_{-}&ia\rho\cos{\vartheta}\sqrt{\Delta}\hat{\mathcal{D}}_{+}&0&0
\end{array} \right). 
\end{equation*}
The operator $\hat{J}$ can be written in a more compact form as follows
\begin{equation} \label{nove}
\hat{J}=\Gamma^{-1}(\Gamma_{(\vartheta)}\mathcal{W}_{(r)}-\Gamma_{(r)}\mathcal{W}_{(\vartheta)}).
\end{equation}
Clearly, a similar expression holds for the Dirac equation in OSC. 
\begin{lemma}
$J=S\hat{J}S^{-1}$ is a symmetry matrix operator for the Dirac operator $\mathcal{O}_{D}$ in the Kerr-Newman metric, i.e.
\begin{equation*}
[\mathcal{O}_{D},J]=0.
\end{equation*}
\end{lemma}
\begin{Proof.}
From (\ref{vdoppio}) we have $\mathcal{O}_{D}=S\Gamma^{-1}\mathcal{W}S^{-1}$, which acts on the spinors $\Psi$ and we consider the operator $J=S\hat{J}S^{-1}$ with $\hat{J}$ given by (\ref{nove}). It holds
\begin{eqnarray*}
\mathcal{O}_{D}J&=&S\Gamma^{-1}\mathcal{W}\Gamma^{-1}(\Gamma_{(\vartheta)}\mathcal{W}_{(r)}-\Gamma_{(r)}\mathcal{W}_{(\vartheta)})S^{-1}\\
J\mathcal{O}_D&=&S\Gamma^{-1}(\Gamma_{(\vartheta)}\mathcal{W}_{(r)}-\Gamma_{(r)}\mathcal{W}_{(\vartheta)})\Gamma^{-1}\mathcal{W}S^{-1}
\end{eqnarray*}
and therefore we obtain
\begin{equation*}
[\mathcal{O}_{D},J]=S\Gamma^{-1}PS^{-1}
\end{equation*}
with
\begin{equation*}
P:=\mathcal{W}\Gamma^{-1}(\Gamma_{(\vartheta)}\mathcal{W}_{(r)}-\Gamma_{(r)}\mathcal{W}_{(\vartheta)})-(\Gamma_{(\vartheta)}\mathcal{W}_{(r)}-\Gamma_{(r)}\mathcal{W}_{(\vartheta)})\Gamma^{-1}\mathcal{W}.
\end{equation*}
Making use of the decomposition $\mathcal{W}=\mathcal{W}_{(r)}+\mathcal{W}_{(\vartheta)}$ and of the commutation relations (\ref{sette}), we obtain
\begin{eqnarray*}
P&=&\mathcal{W}_{(\vartheta)}(\Gamma_{(r)}\Gamma^{-1}+\Gamma^{-1}\Gamma_{(\vartheta)})\mathcal{W}_{(r)}-\mathcal{W}_{(r)}(\Gamma^{-1}\Gamma_{(r)}+\Gamma_{(\vartheta)}\Gamma^{-1})\mathcal{W}_{(\vartheta)}\\
&=&\mathcal{W}_{(\vartheta)}\mathcal{W}_{(r)}-\mathcal{W}_{(r)}\mathcal{W}_{(\vartheta)}=0
\end{eqnarray*}
where in the last line we employed (\ref{4}).
\vspace{-0.2cm}
\begin{equation*}
\hspace{12cm}\square
\end{equation*}
\end{Proof.}
Concerning the physical meaning of $J$, it is interesting to observe that, when $a=0$ the angular eigenfunctions $S_{\pm}(\vartheta)$ can be expressed in terms of the spin-weighted spherical harmonics $Y_{\pm\frac{1}{2}}^{jk}$ (see for instance \cite{New1},\cite{Gold}). According to \cite{Cart1}, we will call $J$ the square root of the total squared angular momentum operator. In \cite{Cart1} it was proved that the separation constant $\lambda$ is the eigenvalue of the square root of the total squared angular momentum for the Dirac equation in any type-D vacuum space-time. Moreover, the occurrence of this operator and therefore the separability of the Dirac equation in the Kerr-Newman metric by means of Chandrasekhar ansatz arises from the presence of a Killing spinor field on the space-time under consideration. For a detailed description of the concept of Killing spinors we refer to \cite{Wal}.

\section{Schr\"{o}dinger form of the Dirac equation in OSC} 
In order to bring the matrix equation 
\begin{equation} \label{vdoppioOSC}
\mathcal{W}^{(OSC)}\hat{\psi}=0
\end{equation}
with $\mathcal{W}^{(OSC)}$ given by (\ref{unostern}) into the form of a Schr\"{o}dinger equation
\begin{equation} \label{Schr}
i\partial_{t}\hat{\psi}=H_{D}\hat{\psi},
\end{equation}
we apply a method similar to that one used in \cite{Fin}. Without risk of confusion we can simplify our notation by omitting to write explicitly the superscript $(OSC)$ attached to the operator $\mathcal{W}$. Starting with (\ref{vdoppioOSC}), we bring the time derivatives on the l.h.s. of the equation and find that
\begin{equation} \label{Sch}
-iT\partial_{t}\hat{\psi}=(\mathcal{W}_{(r,\varphi)}+\mathcal{W}_{(\vartheta,\varphi)})\hat{\psi}
\end{equation} 
with 
\begin{equation*}
T=\left( \begin{array}{cccc}
                            0&0&i\sqrt{\Delta}&-a\sin{\vartheta}\\
                            0&0&-a\sin{\vartheta}&-i\sqrt{\Delta}\\
                            -i\sqrt{\Delta}&-a\sin{\vartheta}&0&0\\
                            -a\sin{\vartheta}&i\sqrt{\Delta}&0&0
         \end{array} \right)
\end{equation*}
and with $\mathcal{W}_{(r,\varphi)}$, $\mathcal{W}_{(\vartheta,\varphi)}$ formally given as in Corollary~\ref{corollary1}, where the operators $\hat{\mathcal{D}}_{\pm}$ and $\hat{\mathcal{L}}_{\pm}$ have been now replaced by 
\begin{eqnarray*}
\tilde{\mathcal{D}}_{\pm}&=&\frac{\partial}{\partial r}\mp\frac{a}{\hat{\Delta}}\frac{\partial}{\partial\varphi}\\
\tilde{\mathcal{L}}_{\pm}&=&\frac{\partial}{\partial \vartheta}+\frac{1}{2}\cot{\vartheta}\mp i\csc{\vartheta}\frac{\partial}{\partial\varphi}.
\end{eqnarray*}
Since det$(T)=\Sigma^2\neq 0$ for $r>0$ and $\vartheta\in[0,\pi]$, $T$ is non singular and we can multiply (\ref{Sch}) on both sides by $-T^{-1}$ to obtain
\begin{equation} \label{ham3}
H_{D}=\frac{\Delta}{\Sigma}\mathcal{S}\cdot(\hat{\mathcal{W}}_{(r,\varphi)}+\hat{\mathcal{W}}_{(\vartheta,\varphi)}),
\end{equation}
where $\sigma_{2}$ is the Pauli matrix and
\begin{eqnarray*}
\mathcal{S}&=&1\hspace{-1mm}\rm{I}_{4}-\frac{a\sin{\vartheta}}{\sqrt{\Delta}}\left( \begin{array}{cc}
       \sigma_{2}&0\\
       0&-\sigma_{2}
           \end{array} \right)\\
\hspace{-1cm}\mathcal{W}_{(r,\varphi)}&=&-\frac{m_{e}r}{\sqrt{\Delta}}\left( \begin{array}{cc}
       0&1\hspace{-1mm}\rm{I}_{2}\\
       1\hspace{-1mm}\rm{I}_{2}&0
           \end{array} \right)+\mbox{diag}(-\mathcal{E}_{-},\mathcal{E}_{+},\mathcal{E}_{+},-\mathcal{E}_{-})\\
\hspace{-1cm}\mathcal{W}_{(\vartheta,\varphi)}&=&\frac{am_{e}\cos{\vartheta}}{\sqrt{\Delta}}\left( \begin{array}{cccc}
                            0&0&i&0\\
                            0&0&0&i\\
                            -i&0&0&0\\
                            0&-i&0&0
         \end{array} \right)+\left( \begin{array}{cccc}
                            0&-\mathcal{M}_{+}&0&0\\
                            -\mathcal{M}_{-}&0&0&0\\
                            0&0&0&\mathcal{M}_{+}\\
                            0&0&\mathcal{M}_{-}&0
         \end{array} \right)
\end{eqnarray*}
with
\begin{equation*}
\mathcal{E}_{\pm}=i\tilde{\mathcal{D}}_{\pm}\quad\mbox{and}\quad\mathcal{M}_{\pm}=\frac{i}{\sqrt{\Delta}}\tilde{\mathcal{L}}_{\pm}
\end{equation*}
such that
\begin{equation*}
\mathcal{E}_{\pm}^{*}=-\mathcal{E}_{\pm}\quad\mbox{and}\quad\mathcal{M}_{\pm}^{*}=-\mathcal{M}_{\mp}.
\end{equation*}
Notice that the formal operator $H_{D}$ acts on the spinors $\hat{\psi}$ on the hypersurfaces $t=$const. Moreover, the matrix $\mathcal{S}$ entering in (\ref{ham3}) is hermitian. For simplicity in the notation we will omit the hat of the wave functions. It can be checked that for every $\psi$, $\phi\in\mathcal{C}_{0}^{\infty}([0,\infty)\times S^2)^{4}$ the Hamiltonian $H_{D}$ is hermitian, i.e. formally self-adjoint with respect to the positive scalar product
\begin{equation} \label{sp1}
\langle \psi|\phi\rangle=\int_{0}^{\infty}dr\int_{-1}^{1}d(\cos{\vartheta})\int_{0}^{2\pi}d\varphi\, (\psi|C\phi)
\end{equation}
with inner product given by
\begin{equation} \label{ip1}
(\psi|C\phi)=\overline{\psi}C\phi,
\end{equation}
where the $\overline{\psi}$ denotes the complex conjugated transposed spinor and
\begin{equation} \label{matriceC}
C=1\hspace{-1mm}\rm{I}_{4}+\frac{a\sin{\vartheta}}{\sqrt{\Delta}}\left( \begin{array}{cc}
       \sigma_{2}&0\\
       0&-\sigma_{2}
           \end{array} \right).
\end{equation}
In order to check the positivity of the scalar product defined by (\ref{sp1}) and (\ref{ip1}) it suffices to show that the matrix $C$ entering in the inner product is positive definite. Indeed, the eigenvalues of $C$ are given by
\begin{equation*}
\Lambda_{1}=\Lambda_{2}=1-\frac{|a|\sin{\vartheta}}{\sqrt{\Delta}}\quad\mbox{and}\quad\Lambda_{3}=\Lambda_{4}=1+\frac{|a|\sin{\vartheta}}{\sqrt{\Delta}}
\end{equation*}
and since the following inequality holds for all $r>0$
\begin{equation*}
\frac{|a|\sin{\vartheta}}{\sqrt{\Delta}}\leq \frac{|a|}{\sqrt{\Delta}}<1,
\end{equation*}
we can conclude that $\Lambda_{i}>0$ for every $i=1,\cdots,4$. In what follows we consider the Hilbert space $\mathcal{H}=\{\mathbb{C}^4,\langle\cdot|\cdot\rangle\}$ consisting of wave functions $\phi:[0,\infty)\times S^{2}\longrightarrow\mathbb{C}^4$ together with the scalar product (\ref{sp1}).

\section{Completeness of Chandrasekhar ansatz} 
We begin with some preliminary observations. Energy, generalized squared angular momentum and the z-component of the total angular momentum form a set of commuting observables $\{H_{D},J^{2},J_{z}\}$. Moreover, the angular system (\ref{angular}) can be brought in the so-called Dirac form \cite{Weidmann}, namely
\begin{equation*}
\mathcal{U}S:=\left( \begin{array}{cc}
       0&1\\
      -1&0
                     \end{array} \right)\frac{dS}{d\vartheta}+\left( \begin{array}{cc}
       -am_{e}\cos{\vartheta}&-\frac{k+\frac{1}{2}}{\sin{\vartheta}}-a\omega\sin{\vartheta}\\
      -\frac{k+\frac{1}{2}}{\sin{\vartheta}}-a\omega\sin{\vartheta}&am_{e}\cos{\vartheta}
                     \end{array} \right)S=\lambda S
\end{equation*}
with $S(\vartheta)=(S_{-}(\vartheta),S_{+}(\vartheta))^{T}$ and $\vartheta\in(0,\pi)$. According to \cite{Araldo}, in $L_{2}((0,\pi))^{2}$ the angular operator $\mathcal{U}$ defined on $D(\mathcal{U})=\mathcal{C}_{0}^{\infty}((0,\pi))^{2}$ is essentially self-adjoint, its spectrum is discrete, non degenerate (i.e. simple) and depends smoothly on $\omega$. Therefore, its eigenvalues can be written as $\lambda_{j}(\omega)$ with $j\in\mathbb{Z}$ and $\lambda_{j}<\lambda_{j+1}$ for every $j\in\mathbb{Z}$. Moreover, the functions $S_{\pm}^{kj}(\vartheta)$ satisfy a generalized Heun equation \cite{Araldo} and become the well-known spin-weighted spherical harmonics by setting $a=0$. Furthermore, the functions $e^{i\left(k+\frac{1}{2}\right)\varphi}$ are eigenfunctions of the z-component of the total angular momentum operator $J_{z}$ with eigenvalues $-\left(k+\frac{1}{2}\right)$ with $k\in\mathbb{Z}$. Hence, we will label the states $\phi$ in the Hilbert space $\mathcal{H}$ by $\phi^{kj}_{\omega}$. In what follows, we consider the free Dirac operator $H_{D}$ given by (\ref{ham3}) in $\mathcal{H}=L_{2}([0,\infty)\times S^{2})^{4}$. From Theorem 1.1 \cite{Theller} the operator $H_{D}$ defined on $\mathcal{C}_{0}^{\infty}([0,\infty)\times S^2)^{4}$ is essentially self-adjoint and has a unique self-adjoint extension on the Sobolev space $W^{1,2}([0,\infty)\times S^2)^{4}$. In addition, the spectrum of $H_{D}$, which we will denote by $\sigma_{D}$, is purely absolutely continuous and given by $\sigma_{D}=(-\infty,m_{e}]\cup[m_{e},+\infty)$. Notice that, since $H_{D}$ possesses a unique self-adjoint extension, when we will derive an integral representation for the Dirac propagator, there will be no need to impose Dirichlet boundary conditions or other boundary conditions on the spinors, as it was done in \cite{Fin}. In preparation of the proof for the completeness of Chandrasekhar ansatz we show in the next Lemma that
\begin{equation} \label{Ydef}
\{Y^{kj}_{\omega}(\vartheta,\varphi)\},\quad Y^{kj}_{\omega}(\vartheta,\varphi)=\frac{1}{\sqrt{2\pi}}\left( \begin{array}{c}
      S_{\omega,-}^{kj}(\vartheta) \\
      S_{\omega,+}^{kj}(\vartheta)
           \end{array} \right)e^{i\left(k+\frac{1}{2}\right)\varphi}
\end{equation}
with $k$, $j\in\mathbb{Z}$ is a complete orthonormal basis of $L_{2}(S^{2})^{2}$.
\begin{lemma}
For every $k$, $j\in\mathbb{Z}$ the set $\{Y^{kj}_{\omega}(\vartheta,\varphi)\}$ with $Y^{kj}_{\omega}(\vartheta,\varphi)$ given by (\ref{Ydef}) is a complete orthonormal basis for $L_{2}(S^{2})^{2}$.
\end{lemma}
\begin{Proof.}
We begin by proving the orthonormality, i.e. we show that for every $k^{'}$, $j^{'}$, $k$, $j\in\mathbb{Z}$ it holds $\langle Y^{kj}_{\omega}|Y^{k^{'}j^{'}}_{\omega}\rangle_{S^{2}}=\delta_{kk^{'}}\delta_{jj^{'}}$, where $\langle\cdot|\cdot\rangle_{S^{2}}$ is the usual scalar product on $S^{2}$.
Making use of (\ref{Ydef}), a direct computation gives
\begin{equation*}
\langle Y^{kj}|Y^{k^{'}j^{'}}\rangle_{S^{2}}=\delta_{kk^{'}}\int_{0}^{\pi}d\vartheta\,\sin{\vartheta}\overline{S}^{kj}(\vartheta)S^{k^{'}j^{'}}(\vartheta),
\end{equation*}
where for simplicity in notation we omitted to write the subscript $\omega$ attached to the angular eigenfunctions. In order to investigate the above integral we need to go back to the angular system (\ref{angular}). It is easy to see that $S^{k^{'}j^{'}}_{+}$, $S^{k^{'}j^{'}}_{-}$ and $\left(S^{kj}_{+}\right)^{*}$, $\left(S^{kj}_{-}\right)^{*}$ satisfy the following two systems of first order linear ODEs, namely
\begin{eqnarray}
+\hat{\mathcal{L}}^{'}_{+}S^{k^{'}j^{'}}_{+}+(\lambda_{j^{'}}-am_{e}\cos\theta)S^{k^{'}j^{'}}_{-}&=&0, \label{laprima}\\
-\hat{\mathcal{L}}^{'}_{-}S^{k^{'}j^{'}}_{-}+(\lambda_{j^{'}}+am_{e}\cos\theta)S^{k^{'}j^{'}}_{+}&=&0 \label{laseconda}
\end{eqnarray}
and
\begin{eqnarray}
+\hat{\mathcal{L}}_{+}\left(S^{kj}_{+}\right)^{*}+(\lambda_{j}-am_{e}\cos\theta)\left(S^{kj}_{-}\right)^{*}&=&0, \label{laterza}\\
-\hat{\mathcal{L}}_{-}\left(S^{kj}_{-}\right)^{*}+(\lambda_{j}+am_{e}\cos\theta)\left(S^{kj}_{+}\right)^{*}&=&0, \label{laquarta}
\end{eqnarray}
where $(\cdot)^{*}$ denotes complex conjugation and
\begin{equation*}
\hspace{-1.9cm}\hat{\mathcal{L}}^{'}_{\pm}=\frac{d}{d\vartheta}+\widetilde{Q}(\vartheta)_{\pm},\quad \widetilde{Q}(\vartheta)_{\pm}=\frac{1}{2}\cot{\vartheta}\pm\widetilde{Q}(\vartheta),\quad\widetilde{Q}(\vartheta)=a\omega\sin{\vartheta}+\left(k^{'}+\frac{1}{2}\right)\csc{\vartheta} .
\end{equation*}
After multiplication of (\ref{laprima}) by $\sin{\vartheta}\left(S^{kj}_{-}\right)^{*}$, of (\ref{laseconda}) by $\sin{\vartheta}\left(S^{kj}_{+}\right)^{*}$, of (\ref{laterza}) by $\sin{\vartheta}S^{k^{'}j^{'}}_{-}$ and of (\ref{laquarta}) by $\sin{\vartheta}S^{k^{'}j^{'}}_{+}$, we consider the equations (\ref{laprima})$\sin{\vartheta}\left(S^{kj}_{-}\right)^{*}-$(\ref{laterza})$\sin{\vartheta}S^{k^{'}j^{'}}_{-}=0$ and (\ref{laseconda})$\sin{\vartheta}\left(S^{kj}_{+}\right)^{*}-$(\ref{laquarta})$\sin{\vartheta}S^{k^{'}j^{'}}_{+}=0$, build their sum and integrate over $\vartheta$ from $0$ to $\pi$. Hence, we obtain
\begin{eqnarray*}
  \lefteqn{2(\lambda_{j^{'}}-\lambda_{j})\int_{0}^{\pi}d\vartheta\,\sin{\vartheta}\overline{S}^{kj}(\vartheta)S^{k^{'}j^{'}}(\vartheta)= }\\
\hspace{-2cm}\int_{0}^{\pi}d\vartheta\,\sin{\vartheta}\left[\left(S^{kj}_{+}\right)^{*}\hat{\mathcal{L}}^{'}_{-}S^{k^{'}j^{'}}_{-}-S^{k^{'}j^{'}}_{+}\hat{\mathcal{L}}_{-}\left(S^{kj}_{-}\right)^{*}-\left(S^{kj}_{-}\right)^{*}\hat{\mathcal{L}}^{'}_{+}S^{k^{'}j^{'}}_{+}+S^{k^{'}j^{'}}_{-}\hat{\mathcal{L}}_{+}\left(S^{kj}_{+}\right)^{*}\right]. 
\end{eqnarray*}
Since the angular eigenfunctions $S_{\pm}(\vartheta)$ even vanish at $\vartheta=0$ and at $\vartheta=\pi$ (see \cite{Araldo}), it can be checked that the following relations hold
\begin{eqnarray*}
+\int_{0}^{\pi}d\vartheta\,\sin{\vartheta}S^{k^{'}j^{'}}_{-}\hat{\mathcal{L}}_{+}\left(S^{kj}_{+}\right)^{*}&=&-\int_{0}^{\pi}d\vartheta\,\sin{\vartheta}\left(S^{kj}_{+}\right)^{*}\hat{\mathcal{L}}_{-}S^{k^{'}j^{'}}_{-}\\
-\int_{0}^{\pi}d\vartheta\,\sin{\vartheta}S^{k^{'}j^{'}}_{+}\hat{\mathcal{L}}_{-}\left(S^{kj}_{-}\right)^{*}&=&+\int_{0}^{\pi}d\vartheta\,\sin{\vartheta}\left(S^{kj}_{-}\right)^{*}\hat{\mathcal{L}}_{+}S^{k^{'}j^{'}}_{+}.
\end{eqnarray*}
By means of the above results we obtain that
\begin{eqnarray*}
\lefteqn{2(\lambda_{j^{'}}-\lambda_{j})\int_{0}^{\pi}d\vartheta\,\sin{\vartheta}\overline{S}^{kj}(\vartheta)S^{k^{'}j^{'}}(\vartheta)= }\\
\hspace{-2cm}\int_{0}^{\pi}d\vartheta\,\sin{\vartheta}\left[\left(S^{kj}_{+}\right)^{*}\left(\hat{\mathcal{L}}^{'}_{-}-\hat{\mathcal{L}}_{-}\right)S^{k^{'}j^{'}}_{-}+\left(S^{kj}_{-}\right)^{*}\left(\hat{\mathcal{L}}_{+}-\hat{\mathcal{L}}^{'}_{+}\right)S^{k^{'}j^{'}}_{+}\right]
\end{eqnarray*}
and this completes the first part of the proof. Regarding the proof of the completeness,  the method we use employs the Projection Theorem (Theorem II.3 \cite{Simon}), i.e. we show that in $L_{2}(S^{2})^{2}$ the only element orthogonal to our orthonormal basis is the zero vector. Without loss of generality let us consider
\begin{equation*}
\widetilde{\varphi}=\frac{1}{\sqrt{2\pi}}\left( \begin{array}{c}
\varphi_{1}(\vartheta)\\
\varphi_{2}(\vartheta)
\end{array} \right)e^{i\left(k+\frac{1}{2}\right)\varphi}\quad\mbox{with}\quad\varphi_{i}\in\mathcal{C}_{0}^{\infty}((0,\pi))\quad\mbox{for}\quad i=1,2.
\end{equation*}
Then we have
\begin{equation*}
\langle\widetilde{\varphi}|Y^{kj}\rangle_{S^{2}}=\int_{-1}^{1}dx\,\left(\varphi_{1}^{*}(x)S_{+}(x)+\varphi_{2}^{*}(x)S_{-}(x)\right),
\end{equation*}
where we made use of the transformation $x=\cos{\vartheta}$. The following estimate holds
\begin{eqnarray*}
\left|\langle\widetilde{\varphi}|Y^{kj}\rangle_{S^{2}}\right|&\leq&\int_{-1}^{1}dx\,\left(\left|\varphi_{1}^{*}(x)\right|\left|S_{+}(x)\right|+\left|\varphi_{2}^{*}(x)\right|\left|S_{-}(x)\right|\right)\\
&\leq&\|\varphi_{1}\|^{2}_{L_{2}}\int_{-1}^{1}dx\,\left|S_{+}(x)\right|^{2}+\|\varphi_{2}\|^{2}_{L_{2}}\int_{-1}^{1}dx\,\left|S_{-}(x)\right|^{2}\\
&\leq&d\int_{-1}^{1}dx\,\left(\left|S_{+}(x)\right|^{2}+\left|S_{-}(x)\right|^{2}\right)=d,
\end{eqnarray*}
where in the second line we used H\"{o}lder inequality, in the third line the orthonormality condition for $S_{\pm}(\vartheta)$ and we defined $d:=\max\{\|\varphi_{1}\|^{2}_{L_{2}},\|\varphi_{2}\|^{2}_{L_{2}}\}$. Clearly, $\left|\langle\widetilde{\varphi}|Y^{kj}\rangle_{S^{2}}\right|\leq 0$ if and only if $d=0$. Since $d=0$ implies $\varphi_{1}=\varphi_{2}=0$ and the scalar product $\langle\cdot|\cdot\rangle_{S^{2}}$ is positive, the proof is completed.  
\vspace{-0.2cm}
\begin{equation*}
\hspace{12cm}\square
\end{equation*}
\end{Proof.} 
In addition from Theorem II.7 in \cite{Simon} it follows that the Hilbert space $L_{2}(S^{2})^{2}$ is separable. We state now the Theorem on the completeness of Chandrasekhar ansatz.
\begin{theorem}
For every $\psi\in\mathcal{C}_{0}^{\infty}([0,+\infty)\times S^{2})^{4}$ it holds
\begin{equation*}
\hspace{-1.5cm}\psi(0,r,\vartheta,\varphi)=\int_{\sigma_{D}}d\omega\,\sum_{k,j\in\mathbb{Z}}\langle\psi^{kj}_{\omega}|\psi_{\omega}\rangle\psi^{kj}_{\omega},\quad \psi^{kj}_{\omega}(r,\vartheta,\varphi)=\left( \begin{array}{c}
               R_{\omega,-}^{kj}(r)Y_{\omega,-}^{kj}(\vartheta,\varphi)\\
               R_{\omega,+}^{kj}(r)Y_{\omega,+}^{kj}(\vartheta,\varphi)\\
               R_{\omega,+}^{kj}(r)Y_{\omega,-}^{kj}(\vartheta,\varphi)\\
               R_{\omega,-}^{kj}(r)Y_{\omega,+}^{kj}(\vartheta,\varphi)
\end{array} \right)
\end{equation*}
with $\sigma_{D}=(-\infty,m_{e}]\cup[m_{e},+\infty)$ and scalar product $\langle\cdot|\cdot\rangle$ defined by (\ref{sp1}).
\end{theorem}
\begin{Proof.}
Let us first recall that Lemma 6.1 implies that the space $C_{0}^{\infty}(S^{2})^{2}$ is spanned by an orthonormal system $\{Y^{kj}_{\omega}(\vartheta,\varphi)\}$. We show now that it is possible to construct isometric operators
\begin{equation*}
\hat{W}_{k,j}:\mathcal{C}_{0}^{\infty}([0,+\infty))^{2}\longrightarrow \mathcal{C}_{0}^{\infty}([0,+\infty)\times S^{2})^{4},
\end{equation*}
such that 
\begin{equation*}
R^{kj}_{\omega}=\left( \begin{array}{c}
               R^{kj}_{\omega,-}(r)\\
               R^{kj}_{\omega,+}(r)
\end{array} \right)\longmapsto A\left( \begin{array}{c}
               R_{\omega,-}^{kj}(r)Y_{\omega,-}^{kj}(\vartheta,\varphi)\\
               R_{\omega,+}^{kj}(r)Y_{\omega,+}^{kj}(\vartheta,\varphi)\\
               R_{\omega,+}^{kj}(r)Y_{\omega,-}^{kj}(\vartheta,\varphi)\\
               R_{\omega,-}^{kj}(r)Y_{\omega,+}^{kj}(\vartheta,\varphi)
\end{array} \right)
\end{equation*}
with $A$ a positive definite hermitian matrix and
\begin{equation*}
Y_{\omega,\pm}^{kj}(\vartheta,\varphi)=\frac{1}{\sqrt{2\pi}}S_{\omega,\pm}^{kj}(\vartheta)e^{i\left(k+\frac{1}{2}\right)\varphi}.
\end{equation*}
Since the angular eigenfunctions $Y^{kj}_{\omega}$ are normalized, we have 
\begin{equation*}
\hspace{-1.9cm}\|R^{kj}_{\omega}\|^{2}_{L_{2}([0,+\infty))^{2}}=\int_{0}^{\infty}dr\int_{S^{2}}d\Omega\, (\psi^{kj}_{\omega}|\psi^{kj}_{\omega})\quad\mbox{with}\quad\psi^{kj}_{\omega}=\left( \begin{array}{c}
               R_{\omega,-}^{kj}Y_{\omega,-}^{kj}\\
               R_{\omega,+}^{kj}Y_{\omega,+}^{kj}\\
               R_{\omega,+}^{kj}Y_{\omega,-}^{kj}\\
               R_{\omega,-}^{kj}Y_{\omega,+}^{kj}
\end{array} \right).
\end{equation*}
Let us define a matrix 
\begin{equation*}
A^{2}:=\frac{\Delta}{\Sigma}\left[1\hspace{-1mm}\rm{I}_{4}+\frac{a\sin{\vartheta}}{\sqrt{\Delta}}\left( \begin{array}{cc}
       \sigma_{2}&0\\
       0&-\sigma_{2}
           \end{array} \right)\right].
\end{equation*}
Clearly, it holds $A^{2}C=1\hspace{-1mm}\rm{I}_{4}$ with $C$ given by (\ref{matriceC}) and $A$ will be a positive definite hermitian matrix. Therefore, taking into account that the matrices $C$ and $A^{2}$ commute, we obtain that
\begin{eqnarray*}
\|R^{kj}\|^{2}_{L_{2}([0,+\infty))^{2}}&=&\int_{0}^{\infty}dr\int_{-1}^{1}d(\cos{\vartheta})\int_{0}^{2\pi}d\varphi\, (\psi^{kj}_{\omega}|CA^{2}\psi^{kj}_{\omega})\\
&=&\langle\psi^{kj}_{\omega}|A^{2}\psi^{kj}_{\omega}\rangle=\langle A\psi^{kj}_{\omega}|A\psi^{kj}_{\omega}\rangle=\langle \hat{W}_{k,j}(R^{kj})|\hat{W}_{k,j}(R^{kj})\rangle\\
&=&\|\hat{W}_{k,j}(R^{kj})\|^{2}_{L_{2}([0,+\infty)\times S^{2})^{4}}.
\end{eqnarray*}
By means of the isometric operators $\hat{W}_{k,j}$ we can now introduce for every $\omega\in\sigma_{D}$ an auxiliary separable Hilbert space $\mathfrak{h}(\omega)$ as follows
\begin{equation*}
\mathfrak{h}(\omega)=\bigoplus_{k,j\in\mathbb{Z}}\mathfrak{h}_{k,j}\quad\mbox{with}\quad\mathfrak{h}_{k,j}=\hat{W}_{k,j}(\mathcal{C}_{0}^{\infty}([0,+\infty))^{2}).
\end{equation*}
Moreover, the expansion theorem (Th.3.7 \cite{Weid}) implies that
\begin{equation*}
\psi_{\omega}=\sum_{k,j\in\mathbb{Z}}\langle\psi^{kj}_{\omega}|\psi_{\omega}\rangle\psi^{kj}_{\omega}.
\end{equation*}
Notice that $\mathfrak{h}(\omega)$ is by no means a subspace of $L_{2}([0,+\infty)\times S^{2})^{4}$ with respect to the spatial measure, since the solutions $R_{\omega,\pm}^{kj}(r)$ of the radial system (\ref{radial}) oscillate asymptotically for $r\to+\infty$. To circumvent this problem we proceed as follows. First we identify the absolutely continuous part of $H_{D}^{(A)}$ with the operator $H_{D}$, since the spectrum $\sigma_{D}$ is purely absolutely continuous. Then, lemma 10 (Ch.1 $\S$3.4 \cite{Jaf}) implies that $H_{D}$ is unitary equivalent to the operator of multiplication by $\omega$ in $L_{2}(\sigma_{D};d\omega)^{4}$. By definition (see Ch.1 $\S$5.1 \cite{Jaf}) the Hilbert space $\mathcal{H}$ is decomposed into the direct integral
\begin{equation} \label{dec}
\mathcal{H}\cong\int_{\sigma_{D}}^{\oplus}d\omega\,\mathfrak{h}(\omega)=:\mathfrak{H},
\end{equation}  
if there is given a unitary mapping $\mathcal{F}$ of the space $\mathcal{H}$ onto $\mathfrak{H}$ but since the operator $H_{D}$ is self-adjoint, the spectral representation theorem (Theorem 7.18, \cite{Weid}) implies the existence of the isomorphism $\mathcal{F}$ and this completes the proof. 
\vspace{-0.2cm}
\begin{equation*}
\hspace{12cm}\square
\end{equation*}
\end{Proof.}
The vector valued function $\psi_{\omega}=(\mathcal{F}\psi)(\omega)$ is called the representative element of the element $\psi\in\mathcal{H}$ in the decomposition (\ref{dec}). The scalar product in $\mathfrak{H}$ will be introduced according to \cite{Jaf}, as follows
\begin{equation*}
\langle \phi|\psi\rangle_{\mathfrak{H}}=\int_{\sigma_{D}}d\omega\,\langle \phi_{\omega}|\psi_{\omega}\rangle,
\end{equation*}
where $\langle\cdot|\cdot\rangle$ denotes the scalar product given by (\ref{sp1}). Furthermore, in $\mathfrak{H}$ we can introduce a norm by means of $\|\cdot\|_{\mathfrak{H}}=(\int_{\sigma_{D}}d\omega\,\|\cdot\|^{2}_{\mathfrak{h}})^{\frac{1}{2}}$.\\
In the next result we give the integral representation of the free Dirac propagator in Minkowski space in oblate spheroidal coordinates. 
\begin{theorem}
For every $\psi\in\mathcal{C}_{0}^{\infty}([0,+\infty)\times S^{2})^{4}$ and $x=(r,\vartheta,\varphi)$ it holds
\begin{equation} \label{rappresentazione}
\hat{\psi}(t,x)=e^{itH_{D}}\psi(0,x)=\int_{\sigma_{D}}d\omega\,e^{i\omega t}\sum_{k,j\in\mathbb{Z}}\psi_{\omega}^{kj}(x)\langle\psi^{kj}_{\omega}|\psi_{\omega}\rangle
\end{equation}
with
\begin{equation*}
\psi^{kj}_{\omega}(x)=\left( \begin{array}{c}
               R_{\omega,-}^{kj}(r)Y_{\omega,-}^{kj}(\vartheta,\varphi)\\
               R_{\omega,+}^{kj}(r)Y_{\omega,+}^{kj}(\vartheta,\varphi)\\
               R_{\omega,+}^{kj}(r)Y_{\omega,-}^{kj}(\vartheta,\varphi)\\
               R_{\omega,-}^{kj}(r)Y_{\omega,+}^{kj}(\vartheta,\varphi)
\end{array} \right),
\end{equation*}
where $\sigma_{D}=(-\infty,m_{e}]\cup[m_{e},+\infty)$ and the scalar product $\langle\cdot|\cdot\rangle$ is defined by (\ref{sp1}).
\end{theorem}
\begin{Proof.}
Since $H_{D}$ is self-adjoint in $\mathcal{H}$ theorem 7.37 \cite{Weid} implies that $\{U(t)=e^{itH_{D}}|t\in\mathbb{R}\}$ is a strongly continuous one-parameter unitary group with $U(t)\psi\in\mathcal{C}^{0}_{\infty}([0,+\infty)\times S^{2})^{4}$ for every $t\in\mathbb{R}$ and every $\psi\in\mathcal{C}^{0}_{\infty}([0,+\infty)\times S^{2})^{4}$. Making use of the version of the spectral theorem as given in \cite{Jaf} (Ch.1 $\S$ 4-5), we have
\begin{equation} \label{prop}
\langle \phi|e^{itH_{D}}\psi\rangle=\int_{\sigma_{D}}d\omega\,e^{i\omega t}\langle \phi_{\omega}|\psi_{\omega}\rangle,
\end{equation}
where $\langle \phi_{\omega}|\psi_{\omega}\rangle$ is given in terms of the resolvent of $H_{D}$ as follows
\begin{equation} \label{resol}
\langle \phi_{\omega}|\psi_{\omega}\rangle=\frac{d}{d\omega}\langle \phi|E(\omega)\psi\rangle=\frac{1}{2\pi i}\lim_{\epsilon\to 0}\langle\phi|[R(\omega-i\epsilon)-R(\omega+i\epsilon)]\psi\rangle
\end{equation}
with $E(\omega)$ the spectral family associated to $H_{D}$. In addition Theorem 1.7 (Ch.10, $\S$1 \cite{Kato}) implies that $\langle \phi_{\omega}|\psi_{\omega}\rangle$ is absolutely continuous in $\omega$, while the unicity of the spectral family $E(\omega)$ and the existence of a spectral family $F(\omega)=\mathcal{F}E(\omega)\mathcal{F}^{-1}$  on $\mathfrak{H}$ follow directly from theorem 7.15 (\cite{Weid}). Making use of the relation (3) in (Ch.1, $\S$4.2 \cite{Jaf}) adapted to our case
\begin{equation*}
R(z)=\int_{-\infty}^{+\infty}\,\frac{1}{\tilde{\omega}-z}dE(\tilde{\omega}),
\end{equation*}
we get
\begin{equation} \label{res1}
\frac{1}{2\pi i}[R(\omega-i\epsilon)-R(\omega+i\epsilon)]=\frac{1}{\pi}\int_{-\infty}^{+\infty}\,\frac{\epsilon}{(\omega-\tilde{\omega})^{2}+\epsilon^{2}}dE(\tilde{\omega}).
\end{equation}
Since the integrand in (\ref{res1}) is bounded and integrable, when we take the limit for $\epsilon\to 0$ in the above expression, we may apply Lebesgue dominated convergence theorem in order to take this limit under the integral sign. Taking into account that
\begin{equation*}
\lim_{\epsilon\to 0}\frac{1}{\pi}\frac{\epsilon}{(\omega-\tilde{\omega})^{2}+\epsilon^{2}}=\delta(\omega-\tilde{\omega}),
\end{equation*} 
we find that
\begin{equation} \label{idi}
\frac{1}{2\pi i}\lim_{\epsilon\to 0}[R(\omega-i\epsilon)-R(\omega+i\epsilon)]=\mbox{Id}.
\end{equation}
Finally, (\ref{idi}) together with (\ref{prop}) and (\ref{resol}) gives the desired result.
\vspace{-0.2cm}
\begin{equation*}
\hspace{12cm}\square
\end{equation*}
\end{Proof.}
In order to construct square integrable wave packets in the coordinate space, we go over in the Minkowski metric in Cartesian coordinates  by means of the inverse of the coordinate transformations (\ref{OSCco}) and apply Theorem 1.8 in \cite{Theller}. Hence, we can choose $\mathcal{H}=L_{2}([0,+\infty)\times S^{2})^{4}$ and normalize the wave packets according to $\langle \psi|\psi\rangle_{\mathcal{H}}=1$.\\
The next Lemma describes locally uniformly in $\omega$ the asymptotics of the solutions of the radial system  
\begin{equation} \label{radialOSC} 
\left( \begin{array}{cc}
     \sqrt{\Delta}\hat{\mathcal{D}}_{-}&-im_{e}r-\lambda\\
     im_{e}r-\lambda&\sqrt{\Delta}\hat{\mathcal{D}}_{+}
           \end{array} \right)\left( \begin{array}{cc}
                                     R_{-} \\
                                     R_{+}
                                     \end{array}\right)=0,
\end{equation}
for $r\to +\infty$, where
\begin{equation*}
\hat{\mathcal{D}}_{\pm}=\frac{d}{dr}\mp i\left[\omega+\frac{\left(k+\frac{1}{2}\right)a}{\Delta}\right],\quad \Delta=r^2+a^2.
\end{equation*}
\begin{lemma}
Every non trivial solution $R$ of (\ref{radialOSC}) for $|\omega|>m_{e}$ behaves asymptotically for $r\to +\infty$ like
\begin{equation*}
R(r)=\left( \begin{array}{cc}
                                     R_{-}(r) \\
                                     R_{+}(r)
                                     \end{array}\right)=\left( \begin{array}{cc}
     \cosh{\Theta}&\sinh{\Theta}\\
     \sinh{\Theta}&\cosh{\Theta}
           \end{array} \right)\left( \begin{array}{cc}
                                     e^{-i\kappa r}[f^{\infty}_{-}+\mathcal{O}(r^{-1})] \\
                                     e^{+i\kappa r}[f^{\infty}_{+}+\mathcal{O}(r^{-1})]
                                     \end{array}\right)
\end{equation*}
with constants $f^{\infty}_{\pm}\neq 0$ and
\begin{equation*}
\Theta=\frac{1}{4}\log{\left(\frac{\omega+m_{e}}{\omega-m_{e}}\right)},\quad\kappa=\epsilon(\omega)\sqrt{\omega^2-m^2_{e}},
\end{equation*}
where $\epsilon(\omega)$ is a sign function such that $\epsilon(\omega)=+1$ for $\omega>m_{e}$ and $\epsilon(\omega)=-1$ for  $\omega<-m_{e}$. 
\end{lemma}
\begin{Proof.}
Let us first rewrite (\ref{radialOSC}) as follows
\begin{equation} \label{ena}
R^{'}(r)=V(r)R(r),\quad V(r)=\left( \begin{array}{cc}
     -i\Omega(r)&\varphi(r)\\
     \varphi^{*}(r)&i\Omega(r)
           \end{array} \right),
\end{equation}
where $*$ denotes complex conjugation and
\begin{equation*}
\Omega(r)=\omega+\frac{\left(k+\frac{1}{2}\right)a}{\Delta},\quad\varphi(r)=\frac{\lambda+im_{e}r}{\sqrt{\Delta}}.
\end{equation*}
By means of the ansatz
\begin{equation*}
R(r)=\hat{A}\tilde{R}(r),\quad \hat{A}=\left( \begin{array}{cc}
     C_{1}&C_{2}\\
     C_{2}&C_{1}
           \end{array} \right),\quad \tilde{R}(r)=\left( \begin{array}{cc}
                                     \tilde{R}_{-}(r) \\
                                     \tilde{R}_{+}(r)
                                     \end{array}\right)
\end{equation*}
with constants $C_{1}$ and $C_{2}$ such that $C_{1}^{2}-C_{2}^{2}\neq 0$, (\ref{ena}) becomes
\begin{equation*}
\tilde{R}^{'}(r)=T(r)\tilde{R}(r),\quad T(r)=\hat{A}^{-1}V(r)\hat{A}=\left( \begin{array}{cc}
     -T_{11}(r)&-T_{12}(r)\\
     T_{12}(r)&T_{11}(r)
           \end{array} \right),
\end{equation*}
where
\begin{eqnarray*}
T_{11}(r)&=&\frac{i(C_{1}^{2}+C_{2}^{2})\Omega(r)+C_{1}C_{2}(\varphi^{*}(r)-\varphi(r))}{C_{1}^{2}-C_{2}^{2}}\\
T_{12}(r)&=&\frac{2iC_{1}C_{2}\Omega(r)+C_{1}^{2}\varphi^{*}(r)-C_{2}^{2}\varphi(r)}{C_{1}^{2}-C_{2}^{2}}.
\end{eqnarray*}
We fix now $C_{1}$ and $C_{2}$ by requiring that
\begin{equation} \label{condizia}
\lim_{r\to +\infty}T_{11}(r)=i\kappa, \quad C_{1}^{2}-C_{2}^{2}=1.
\end{equation}
Taking into account that asymptotically for $r\to+\infty$, it results that
\begin{equation*}
T_{11}(r)=\frac{i(C_{1}^{2}+C_{2}^{2})\omega-2im_{e}C_{1}C_{2}}{C_{1}^{2}-C_{2}^{2}}+\mathcal{O}\left(\frac{1}{r^2}\right),
\end{equation*}
condition (\ref{condizia}) becomes
\begin{equation*}
(C_{1}^{2}+C_{2}^{2})\omega-2m_{e}C_{1}C_{2}-\kappa=0.
\end{equation*}
A simple calculation gives
\begin{equation} \label{forme}
C_{1}=\cosh{\Theta},\quad C_{2}=\sinh{\Theta},\quad\Theta=\frac{1}{4}\log{\left(\frac{\omega+m_{e}}{\omega-m_{e}}\right)}.
\end{equation}
With the help of (\ref{forme}) $T_{11}$ and $T_{12}$ can be rewritten as follows
\begin{eqnarray*}
T_{11}(r)&=&\frac{i}{\kappa}\left(\omega\Omega(r)-m^{2}_{e}\frac{r}{\sqrt{\Delta}}\right)\\
T_{12}(r)&=&\frac{\lambda}{\sqrt{\Delta}}+i\frac{m_{e}}{\kappa}\left(\Omega(r)-\omega\frac{r}{\sqrt{\Delta}}\right).
\end{eqnarray*}
Let us now introduce the decomposition $T(r)=A(r)+B(r)$ with
\begin{equation*}
A(r)=\left( \begin{array}{cc}
     -\frac{i}{\kappa}\left(\omega^{2}-m^{2}_{e}\frac{r}{\sqrt{\Delta}}\right)&-\frac{\lambda}{\sqrt{\Delta}}-im_{e}\frac{\omega}{\kappa}\left(1-\frac{r}{\sqrt{\Delta}}\right)\\
  \frac{\lambda}{\sqrt{\Delta}}+im_{e}\frac{\omega}{\kappa}\left(1-\frac{r}{\sqrt{\Delta}}\right)&\frac{i}{\kappa}\left(\omega^{2}-m^{2}_{e}\frac{r}{\sqrt{\Delta}}\right)
           \end{array} \right)
\end{equation*}
and
\begin{equation*}
B(r)=\frac{i}{\kappa}\frac{\left(k+\frac{1}{2}\right)a}{\Delta}\left( \begin{array}{cc}
     -1&-m_{e}\\
     m_{e}&1
           \end{array} \right).
\end{equation*}
Since $A(r)$ and $B(r)$ are continuously differentiable on $[0,+\infty)$,
\begin{equation*}
\int_{0}^{+\infty}\, dr|A^{'}(r)|<\infty,\quad \int_{0}^{+\infty}\, dr|B(r)|<\infty
\end{equation*}
and the eigenvalues $\lambda_{1}$ and $\lambda_{2}$ of the matrix $A_{0}$ defined by
\begin{equation*}
A_{0}:=\lim_{r\to+\infty}A(r)=\left( \begin{array}{cc}
     -i\kappa&0\\
      0&i\kappa
           \end{array} \right)
\end{equation*}
are simple, then Theorem 11 in \cite{Coppel}, which is actually a simplified version of a theorem due to Levinson \cite{Lev}, implies that the equation $\tilde{R}^{'}(r)=T(r)\tilde{R}(r)$ has solutions such that for $r\to +\infty$
\begin{equation*}
\tilde{R}(r)=\left( \begin{array}{cc}
                                     e^{-i\kappa r}[f^{\infty}_{-}+\mathcal{O}(r^{-1})] \\
                                     e^{+i\kappa r}[f^{\infty}_{+}+\mathcal{O}(r^{-1})]
                                     \end{array}\right),
\end{equation*}
where $f_{-}^{\infty}$ and $f_{+}^{\infty}$ are the non zero components of the eigenvectors $\xi_{-}$ and $\xi_{+}$ belonging to the eigenvalues $-i\kappa$ and $+i\kappa$ of the matrix $A_{0}$, respectively.
\vspace{-0.2cm}
\begin{equation*}
\hspace{12cm}\square
\end{equation*}
\end{Proof.}
\ack 
The work of D.B. was supported by EU grant HPRN-CT-2002-00277. One of us (D.B.) wants to thank the McGill University, Montreal for hospitality, when part of this work was developed.


\section*{References}
\begin{thebibliography}{10}
\bibitem{Haf1} H\"{a}fner D, 2003 {Dissertationes Math. 421, 102}
\bibitem{Da} Daud$\acute{\mbox{e}}$ T, 2004 {\it Sur la th$\acute{\mbox{e}}$orie de la diffusion pour des champs de Dirac dans divers espaces-temps de la relativit$\acute{\mbox{e}}$ g$\acute{\mbox{e}}$n$\acute{\mbox{e}}$rale} (PhD Thesis, University of Bordoux I)
\bibitem{New} Newman E T, Penrose R 1962 {\JMP 3, 566}
\bibitem{Page} Page D 1976 {\PR D 14, 1509}
\bibitem{Roger} Penrose R, Rindler W 1986 {\it Spinors and Space-Time} (Cambridge University Press) Vol.1
\bibitem{Cart} Carter B 1972 {\it Black holes/Les astres occlus} (Ecole d'ete Phys. Theor., Les Houches)
\bibitem{Flamm} Flammer C 1957 {\it Spheroidal Wave Functions} (Stanford University Press, Stanford, California)
\bibitem{Boy} Boyer R H, Lindquist R W 1967 {\JMP 8, No.2, 265}
\bibitem{Kinn} Kinnersley W 1969 {\JMP 10, 1195}
\bibitem{Bose} Bose S K 1975 {\JMP 16, 772}
\bibitem{Goldberg} Goldberg J, Sachs R 1962 {\it Acta Phys. Polon.} 22, 13
\bibitem{Mis} Misner C W, Thorne K S, Wheeler J A 1973 {\it Gravitation} (San Francisco: Freman W H) p 898
\bibitem{Cart1} Carter B, McLenaghan R G 1979 {\PR D 19, No 4, 1093}
\bibitem{Chandra} Chandrasekhar F R S 1976 {\PRS Lond. A 349, 571}
\bibitem{New1} Newman E T, Penrose R 1966 {\JMP 7, 863}
\bibitem{Gold} Goldberg J N, Macfarlane A J, Newman E T, Sudarsan E C G 1967 {\JMP 8, 2155}
\bibitem{Wal} Walker M, Penrose R 1970 {\it Commun. Math. Phys.} 18, 265 
\bibitem{Fin} Finster F, Kamran N, Smoller J, Yau S-T 2003 {\it Adv. Theor. Math. Phys.} 7, 25
\bibitem{Weidmann} Weidmann J 1987 {\it Spectral Theory of Ordinary Differential Operators} (Springer-Verlag Berlin, Heidelber, New York, Lecture Notes in Mathematics, 1258)
\bibitem{Araldo} Batic D, Schmid H, Winklmeier M 2005 {\JMP 46, 012504}
\bibitem{Theller} Theller B 1992 {\it The Dirac Equation} (Springer-Verlag Berlin, Heidelberg, New York, Texts and Monographs in Physics)
\bibitem{Simon} Reed M, Simon B 1980 {\it Methods of Modern Mathematical Physics} (Academic Press Inc., Vol.I) 
\bibitem{Jaf} Jafaev D R 1992 {\it Mathematical Scattering Theory} (Translations of Mathematical Monographs, Vol.105, American Mathematical Society, Providence, Rhode Island)
\bibitem{Weid} Weidmann J 1980 {\it Linear Operators in Hilbert Spaces} (Springer-Verlag New York, Graduate Texts in Mathematics, 68)
\bibitem{Kato} Kato T 1966 {\it Perturbation Theory for linear operators} (Springer Verlag Berlin, Heidelberg, New York, Die Grundlehren der mathematischen Wissenschaften, Band 132)
\bibitem{Coppel} Coppel W A 1965 {\it Stability and Asymptotic Behavior of Differential Equations} (Heath Mathematical Monographs)
\bibitem{Lev} Levinson N 1948 {\it Duke Math. J.} 15, 111
\end{thebibliography}
\end{document}